\documentclass[trackchanges, twocolumn]{aastex701}

\usepackage{comment}
\usepackage{xcolor}

\begin{document}

\title{PRISMS/U13151. A Glimpse into the Earliest Phases of Galaxy Formation through an Ultra-Faint System $\approx$ 500 Myr after the Big Bang}

\author[orcid=0000-0002-5104-8245,sname='Pierluigi Rinaldi']{Pierluigi Rinaldi}
\altaffiliation{Cosmic Frontier Center Prize Fellow}
\affiliation{Space Telescope Science Institute, 3700 San Martin Drive, Baltimore, Maryland 21218, USA}
\affiliation{Department of Astronomy, The University of Texas at Austin, Austin, TX 78712, USA}
\affiliation{Cosmic Frontier Center, The University of Texas at Austin, Austin, TX 78712, USA}
\email[show]{prinaldi@utexas.edu} 
\correspondingauthor{Pierluigi Rinaldi}

\author[orcid=0000-0002-7093-1877,sname='Javier \'Alvarez-M\'arquez']{Javier \'Alvarez-M\'arquez}
\email[]{} 
\affiliation{Centro de Astrobiolog\'ia (CAB), CSIC-INTA, Cra. de Ajalvir Km.4, 28850 Torrej\'on de Ardoz, Madrid, Spain}

\author[orcid=0000-0003-2119-277X, sname='Alejandro Crespo G\'omez']{Alejandro Crespo G\'omez}
\email[]{} 
\affiliation{Space Telescope Science Institute, 3700 San Martin Drive, Baltimore, Maryland 21218, USA}

\author[orcid=0000-0002-9090-4227, sname='Luis Colina']{Luis Colina}
\email[]{} 
\affiliation{Centro de Astrobiolog\'ia (CAB), CSIC-INTA, Cra. de Ajalvir Km.4, 28850 Torrej\'on de Ardoz, Madrid, Spain}

\author[orcid=0000-0002-6719-380X, sname='Stefano Carniani']{Stefano Carniani}
\email[]{} 
\affiliation{Scuola Normale Superiore, Piazza dei Cavalieri 7, I-56126 Pisa, Italy}

\author[orcid=0000-0002-7612-0469, sname='Sarah Kendrew']{Sarah Kendrew}
\email[]{}
\affiliation{AURA for the European Space Agency (ESA), Space Telescope Science Institute, 3700 San Martin Dr., Baltimore, MD 21218, USA}

\author[orcid=0000-0001-8386-3546,sname='Edoardo Iani']{Edoardo Iani}
\email[]{} 
\affiliation{Institute of Science and Technology Austria (ISTA), Am Campus 1, 3400 Klosterneuburg, Austria}

\author[orcid=0000-0002-8192-8091, sname='Angela Adamo']{Angela Adamo}
\email[]{}
\affiliation{The Oskar Klein Centre, Department of Astronomy, Stockholm University, AlbaNova, SE-106 91 Stockholm, Sweden}

\author[orcid=0000-0002-5258-8761, sname='Abdurrouf']{Abdurro'uf}
\email[]{}
\affiliation{Department of Astronomy, Indiana University, 727 East Third Street, Bloomington, IN 47405, USA}

\author[orcid=0000-0002-8053-8040,sname='Marianna Annyziatella']{Marianna Annuziatella}
\email[]{}
\affiliation{INAF – IASF Milano, via A. Corti 12, I-20133 Milano, Italy}

\author[orcid=0000-0002-8909-8782, sname='Stacey Alberts']{Stacey Alberts}
\email[]{}
\affiliation{AURA for the European Space Agency (ESA), Space Telescope Science Institute, 3700 San Martin Dr., Baltimore, MD 21218, USA}

\author[orcid=0000-0002-3952-8588, sname='Leindert A. Boogaard']{Leindert A. Boogaard}
\email[]{}
\affiliation{Leiden Observatory, Leiden University, PO Box 9513, NL-2300 RA Leiden, The Netherlands}

\author[orcid=0000-0001-8183-1460, sname='Karina I. Caputi']{Karina I. Caputi}
\email[]{}
\affiliation{Kapteyn Astronomical Institute, University of Groningen, P.O. Box 800, 9700AV Groningen, The Netherlands}

\author[orcid=0000-0001-9875-8263, sname='Marco Castellano']{Marco Castellano}
\email[]{}
\affiliation{INAF - Osservatorio Astronomico di Roma, via di Frascati 33, 00078 Monte Porzio Catone, Italy}

\author[orcid=0009-0001-8880-944X, sname='Michele Catone']{Michele Catone}
\email[]{}
\affiliation{Dipartimento di Fisica e Astronomia "G. Galilei", Università di Padova, Vicolo dell’Osservatorio 3, 35131 Padova, Italy}
\affiliation{INAF, Osservatorio Astronomico di Padova, Vicolo dell’Osservatorio 5, 35122 Padova, Italy}

\author[orcid=0000-0001-8442-1846, sname='Rui Marques-Chaves']{Rui Marques-Chaves}
\email[]{}
\affiliation{D\'epartement d'Astronomie, Universit\'e de Gen\`eve, Chemin Pegasi 51, CH-1290 Versoix, Switzerland}

\author[orcid=0009-0000-0413-5699, sname='Ryan Cooper']{Ryan Cooper}
\email[]{}
\affiliation{Kapteyn Astronomical Institute, University of Groningen, P.O. Box 800, 9700AV Groningen, The Netherlands}

\author[orcid=0000-0001-7201-5066 , sname='Seiji Fujimoto']{Seiji Fujimoto}
\email[]{}
\affiliation{David A. Dunlap Department of Astronomy and Astrophysics, University of Toronto, 50 St. George Street, Toronto, Ontario, Canada}
\affiliation{Dunlap Institute for Astronomy and Astrophysics, 50 St. George Street, Toronto, Ontario, M5S 3H4, Canada}

\author[orcid=0000-0001-9626-9642, sname='Mahmoud Hamed']{Mahmoud Hamed}
\email[]{}
\affiliation{Centro de Astrobiolog\'ia (CAB), CSIC-INTA, Ctra. de Ajalvir km 4, Torrej\'on de Ardoz, E-28850, Madrid, Spain}

\author[orcid=0000-0002-6047-430X, sname='Yuichi Harikane']{Yuichi Harikane}
\email[]{}
\affiliation{Institute for Cosmic Ray Research, The University of Tokyo, 5-1-5 Kashiwanoha, Kashiwa, Chiba 277-8582, Japan}

\author[orcid=0000-0003-0129-2079, sname='Santosh Harish']{Santosh Harish}
\email[]{}
\affiliation{Space Telescope Science Institute, 3700 San Martin Drive, Baltimore, Maryland 21218, USA}

\author[orcid=0000-0002-6963-4322, sname='Takuya Hashimoto']{Takuya Hashimoto}
\email[]{}
\affiliation{Division of Physics, Faculty of Pure and Applied Sciences, University of Tsukuba, Tsukuba, Ibaraki 305-8571, Japan}
\affiliation{Tomonaga Center for the History of the Universe (TCHoU), Faculty of Pure and Applied Sciences, University of Tsukuba, Tsukuba, Ibaraki 305-8571, Japan}

\author[orcid=0000-0003-4337-6211, sname='Jakob M. Helton']{Jakob M. Helton}
\email[]{}
\affiliation{Department of Astronomy and Astrophysics, The Pennsylvania State University, University Park, PA 16802, USA}

\author[orcid=0000-0003-4512-8705,sname='Tiger Y. -Y. Hsiao']{Tiger Y. -Y. Hsiao}
\altaffiliation{Cosmic Frontier Center Prize Fellow}
\email[]{}
\affiliation{Department of Astronomy, The University of Texas at Austin, Austin, TX 78712, USA}
\affiliation{Cosmic Frontier Center, The University of Texas at Austin, Austin, TX 78712, USA}

\author[orcid=0000-0002-5588-9156,sname='Vasily Kokorev']{Vasily Kokorev}
\altaffiliation{Hubble Fellow}
\email[]{}
\affiliation{Department of Astronomy, The University of Texas at Austin, Austin, TX 78712, USA}
\affiliation{Cosmic Frontier Center, The University of Texas at Austin, Austin, TX 78712, USA}

\author[orcid=0000-0002-0690-8824, sname='Alvaro Labiano']{\'Alvaro Labiano}
\email[]{} 
\affiliation{Telespazio UK for the European Space Agency, ESAC, Camino Bajo del Castillo s/n, 28692 Villanueva de la Ca\~nada, Spain}

\author[orcid=0000-0003-0470-8754, sname='Jens Melinder']{Jens Melinder}
\email[]{}
\affiliation{Department of Astronomy, Stockholm University, Oscar Klein Centre, AlbaNova University Centre, 106 91 Stockholm, Sweden}

\author[orcid=0000-0001-7300-9450, sname='Ikki Mitsuhashi']{Ikki Mitsuhashi}
\email[]{}
\affiliation{Department for Astrophysical \& Planetary Science, University of Colorado, Boulder, CO 80309, USA}

\author[orcid=0000-0002-3005-1349, sname='G\"oran \"Ostlin']{G\"oran \"Ostlin}
\email[]{}
\affiliation{Department of Astronomy, Stockholm University, Oscar Klein Centre, AlbaNova University Centre, 106 91 Stockholm, Sweden}

\author[orcid=0000-0002-3005-1349, sname='Lorenzo Napolitano']{Lorenzo Napolitano}
\email[]{}
\affiliation{INAF - Osservatorio Astronomico di Roma, via di Frascati 33, 00078 Monte Porzio Catone, Italy}

\author[orcid=0000-0002-0939-9156, sname='Borja P\'erez-D\'iaz']{Borja P\'erez-D\'iaz}
\email[]{}
\affiliation{INAF - Osservatorio Astronomico di Roma, via di Frascati 33, 00078 Monte Porzio Catone, Italy}

\author[orcid=0000-0003-4528-5639, sname='Pablo G. P{\'e}rez-Gonz{\'a}lez']{Pablo G. P{\'e}rez-Gonz{\'a}lez}
\email[]{}
\affiliation{Centro de Astrobiolog\'ia (CAB), CSIC-INTA, Cra. de Ajalvir Km.4, 28850 Torrej\'on de Ardoz, Madrid, Spain}

\author[orcid=0009-0005-4109-161X, sname='Carlota Prieto Jim\'enez']{Carlota Prieto Jim\'enez}
\email[]{}
\affiliation{Centro de Astrobiolog\'ia (CAB), CSIC-INTA, Cra. de Ajalvir Km.4, 28850 Torrej\'on de Ardoz, Madrid, Spain}
\affiliation{Departamento de F\'{i}sica de la Tierra y Astrof\'{i}sica, Facultad de Ciencias F\'{i}sicas, Universidad Complutense de Madrid, E-28040, Madrid, Spain}

\author[orcid=0000-0001-7095-7543, sname='Giulia Rodighiero']{Giulia Rodighiero}
\email[]{}
\affiliation{Dipartimento di Fisica e Astronomia "G. Galilei", Università di Padova, Vicolo dell’Osservatorio 3, 35131 Padova, Italy}
\affiliation{INAF, Osservatorio Astronomico di Padova, Vicolo dell’Osservatorio 5, 35122 Padova, Italy}

\author[orcid=0000-0002-9334-8705, sname='Paola Santini']{Paola Santini}
\email[]{}
\affiliation{INAF - Osservatorio Astronomico di Roma, via di Frascati 33, 00078 Monte Porzio Catone, Italy}

\author[orcid=0009-0005-7278-1541, sname='Ana Varo-O''Ferrall']{Ana Varo-O'Ferrall}
\email[]{}
\affiliation{Centro de Astrobiolog\'ia (CAB), CSIC-INTA, Cra. de Ajalvir Km.4, 28850 Torrej\'on de Ardoz, Madrid, Spain}
\affiliation{Departamento de F\'{i}sica de la Tierra y Astrof\'{i}sica, Facultad de Ciencias F\'{i}sicas, Universidad Complutense de Madrid, E-28040, Madrid, Spain}

\author[orcid=0000-0002-7051-1100, sname='Jorge A. Zavala']{Jorge A. Zavala}
\email[]{}
\affiliation{University of Massachusetts Amherst, 710 North Pleasant Street, Amherst, MA 01003-9305, USA}

\author[orcid=0000-0003-3307-7525, sname='Yongda Zhu']{Yongda Zhu}
\email[]{}
\affiliation{Steward Observatory, University of Arizona, 933 North Cherry Avenue, Tucson, AZ 85721, USA}


\begin{abstract}

Detailed UV-to-optical studies of galaxies at $z\gtrsim10$ have largely focused on bright systems ($M_{\rm UV}\lesssim-20$ mag), leaving intrinsically faint galaxies virtually unexplored to date. Combining NIRCam, NIRSpec, and MIRI imaging and spectroscopy, we investigate UNCOVER-ABELL2744-13151 (U13151), an ultra-faint galaxy at $z=9.7995$ ($M_{\rm UV}=-17.44$ mag), magnified by Abell 2744 ($\mu=12.8$). U13151 is very blue ($\beta\approx-2.64$), ultra-low mass ($\log_{10}(M_\bigstar/M_\odot)\approx6.68$), young ($\approx5.5$ Myr), metal-poor ($Z\approx3$--$6\%\,Z_\odot$), and compact ($R_{\rm e}\approx122$ pc). It exhibits negligible dust attenuation, extreme ionization conditions, and intense star formation (${\rm sSFR}_{\rm H\alpha}\approx94\,{\rm Gyr}^{-1}$, or equivalently $M_\bigstar/\rm SFR_{H\alpha}\approx10.6$ Myr). Despite similarities with much brighter galaxies, its low C/O, approximately solar Ne/O, and undetected nitrogen emission suggest enrichment dominated by core-collapse supernovae, in stark contrast with similar nitrogen-enhanced systems. U13151 might be undergoing its first major episode of star formation, assembling possibly most (if not all) of its $M_\bigstar$ through five unresolved clumps. Its high ionizing photon production efficiency ($\log_{10}(\xi_{\rm ion}/({\rm Hz\,erg^{-1}}))\approx25.64$) suggests, through a simple toy model, that similarly short-lived episodes could contribute significantly to the ionizing photon budget of low-mass galaxies during Cosmic Reionization. We explore different evolutionary scenarios, from U13151-like galaxies potentially representing the progenitors of Blue Monsters at higher redshifts to the possibility that such systems could be the early building blocks of present-day Milky Way-like galaxies. Our results highlight the fundamental role of MIRI in studying galaxies in the early Universe and the urgent need to investigate the faintest, lowest-mass systems to shed light on the fundamental physics driving the earliest phases of galaxy formation.

\end{abstract}



\section{Introduction}
One of the primary motivations behind JWST (\citealt{gardner_james_2023}) was to look so far back in cosmic time that we could be able to capture the light from the first generations of stars and witness directly the earliest stages of galaxy formation and the processes that ultimately shaped the Universe as we observe it today. 

Remarkably, in its first years of operations, JWST has already begun to fulfill this promise, with spectroscopically confirmed galaxies now reaching even $z\gtrsim14$ (\citealt{carniani_spectroscopic_2024, naidu_cosmic_2026}), pushing the redshift frontier to epochs that would have been almost unthinkable before its launch. Nowadays, photometric candidates have even been reported up to $z\approx25$ (e.g., \citealt{castellano_pushing_2025, perez-gonzalez_rise_2025}).

Perhaps even more strikingly, JWST has revealed that early galaxies are both remarkably abundant and unexpectedly luminous (e.g., \citealt{naidu_two_2022, arrabal_haro_confirmation_2023, bunker_jades_2023, curtis-lake_spectroscopic_2023, carniani_spectroscopic_2024, napolitano_seven_2025}). Early observations with JWST showed that galaxy formation within the first $\approx500$ Myr of cosmic history ($z>9$) appears to have been more efficient than predicted by empirical extrapolations of pre-JWST samples and by many state-of-the-art models and simulations (e.g., \citealt{castellano_early_2022, perez-gonzalez_life_2023, finkelstein_complete_2024}). 

These luminous early galaxies have been labeled “Blue Monsters” (\citealt{ferrara_stunning_2023}) because of their exceptionally blue rest-frame ultraviolet (UV) continua, with UV slopes bluer than $\approx-2.5$ (e.g., \citealt{cullen_ultraviolet_2024, hainline_cosmos_2024, morales_rest-frame_2024}). Usually, such blue colors are associated with young, metal-poor stellar populations with little-to-no dust. Recent analyses have revealed that these Blue Monsters are not only bright ($M_{\rm UV}\lesssim-20$), but also extremely compact ($R_{\rm e}\approx30\text{--}500$ pc; \citealt{ono_morphological_2025}) and quite massive for such high redshifts ($M_\bigstar\approx10^9\,M_\odot$; e.g., \citealt{roberts-borsani_jwst_2026}). Altogether, these initial results have challenged our conventional understanding based on pre-JWST models. Indeed, several scenarios have been proposed to explain their abundance and existence. These include radiation-driven dusty outflows that reduce dust attenuation and enhance the emergent UV luminosity (e.g., \citealt{ferrara_eventful_2024}); feedback-free or weak-feedback starbursts with temporarily elevated star-formation efficiencies (\citealt{dekel_efficient_2023, yung_cdm_2025, somerville_density-modulated_2025}); and stochastic star-formation histories, in which short-lived bursts scatter low-mass galaxies into the bright end of the UV luminosity function (e.g., \citealt{mason_brightest_2023, munoz_breaking_2023, pallottini_stochastic_2023, gelli_impact_2024, mitsuhashi_uncovermegascience_2026}).

Although these UV-bright galaxies are rapidly transforming our view of the early Universe, they likely provide only a partial and biased picture of early galaxy formation, tracing the most readily detectable tail of the population and perhaps a particular evolutionary phase. A more complete picture {\it must} include intrinsically faint galaxies, which are expected to be far more numerous and may represent the fundamental building blocks of the first cosmic structures. Indeed, the UV luminosity function at high redshift implies that galaxies with $M_{\rm UV}\approx-17$ mag are typically $\approx1.5$ dex, or $\approx30$ times, more abundant than galaxies with $M_{\rm UV}\approx-20$ mag (e.g., \citealt{perez-gonzalez_life_2023}). Yet, their low luminosities make them exceptionally difficult to identify. For instance,  \citet{hainline_jwst_2026} compiled more than 2000 photometric candidates at $z\gtrsim8$ in GOODS-North and GOODS-South using the latest JADES DR5 catalogs (\citealt{robertson_jwst_2026, johnson_jwst_2026}), but galaxies with $M_{\rm UV}\gtrsim-18$ mag at $z\gtrsim9.5\text{--}10$ still account for only $\lesssim10\text{--}15\%$ of the sample. Detection of this faint population is not sufficient though, as their detailed characterization for, e.g., metallicities, dust content, ionization conditions, and star-formation histories (SFHs), requires ultra-deep spectroscopy that remains {\it extremely} demanding in blank fields. Such measurements were possible mainly for brighter systems at similar epochs (e.g., \citealt{bunker_jades_2023, carniani_shining_2024, zavala_luminous_2025, calabro_evidence_2024, carniani_intense_2026, helton_ionizing_2026, naidu_cosmic_2026}), and even in that case observations were expensive. For intrinsically faint galaxies, such a detailed spectroscopic characterization is challenging in blank fields, even with JWST.

Nonetheless, gravitational lensing provides a powerful complementary route for studying intrinsically faint galaxies ($M_{\rm UV}\gtrsim-18$ mag), magnifying their fluxes and enabling direct constraints on their physical properties. For example, \citet{nakajima_ultra-faint_2026} reported that LAP1-B (\citealt{vanzella_extremely_2023}), an ultra-faint galaxy ($M_{\rm UV}>-10.4$ mag) at $z_{\rm spec}=6.62$ behind MACS J0416, has an exceptionally low gas-phase metallicity of $12+\log({\rm O/H})=6.31^{+0.15}_{-0.23}$, corresponding to only $(4.2\pm1.8)\times10^{-3}\,Z_\odot$. Similarly, \citet{korber_glimpse-d_2026} used NIRSpec/G395M to study galaxies down to $M_{\rm UV}\approx-14$ behind Abell S1063 (\citealt{atek_jwsts_2025}) at $z\approx6\text{--}9$, finding a rapid decline in metallicity toward fainter UV luminosities, consistent with earlier JWST results (\citealt{roberts-borsani_nature_2023}).

Beyond their integrated properties, JWST is also resolving the internal structure of faint high-redshift galaxies on very small scales (e.g., tens of parsecs). Some examples of such systems are, for instance, the Cosmic Gems Arc, Firefly Sparkle, Misty Moons, Cosmic Archipelago, Cosmic Grapes,  BulletArc-z11, and Cosmic Spear (\citealt{adamo_bound_2024, bradley_unveiling_2025, messa_jwst_2026, mowla_formation_2024, nakane_venus_2025, messa_anatomy_2025, bolamperti_cosmic_2026, fujimoto_primordial_2025, abdurrouf_spatially_2025, bradac_star_2025}). These initial studies suggested that, at least in these these systems, early stellar-mass assembly could have mainly proceeded through clumps. Together, these discoveries show that JWST is beginning to {\it expose} galaxy formation on its fundamental physical scales, allowing us to investigate how the earliest stellar systems assembled into the galaxies observed in the present-day Universe. Yet, for very faint systems at the highest redshifts studied so far, rest-frame optical emission lines (e.g., H$\beta$, [O\,{\sc iii}], H$\alpha$) often fall beyond the NIRSpec wavelength coverage, limiting direct constraints their physical properties.

In this study, we take a step forward by presenting the first detailed rest-frame UV-to-optical analysis of UNCOVER-ABELL2744-13151 (U13151 hereafter), an intrinsically ultra-faint galaxy ($M_{\rm UV}\approx-17.44$ mag\footnote{Corrected for lensing effects.}) observed when the Universe was only $\approx500$~Myr old ($z\approx9.8$). U13151 is part of the sample of high-$z$ galaxies targeted by {\it The PRImordial galaxy Survey with MIRI Spectroscopy} (PRISMS) program (PID 8051; PIs: J. \'{A}lvarez-M\'{a}rquez \& L. Colina, see also \citealt{alvarez-marquez_prisms_2026, marques-chaves_prisms_2026, perez-diaz_prisms_2026}). U13151 was first identified by \citet{roberts-borsani_nature_2023} and later by \citet{fujimoto_uncover_2024} behind the Abell 2744 lensing cluster as part of the UNCOVER survey (\citealt{bezanson_jwst_2024}) and is strongly magnified, with $\mu=12.8\pm0.7$ (\citealt{furtak_uncovering_2023, price_uncover_2024}). 

Here, we combine NIRSpec and NIRCam with the newly acquired MIRI Low Resolution Spectrometer (LRS) and MIRI imaging data, the latter extending our coverage into the rest-frame optical and enabling us to trace H$\alpha$ in such a system. This unique combination of datasets provide us, for the first time, with a glimpse into the fundamental physics governing the earliest stages of galaxy formation in the early Universe.

\vspace{2mm}
This paper is organized as follows. In Section \ref{sec2}, we provide an overview of the data used in this work for our analysis, with a brief description of data reduction for NIRSpec and MIRI (imaging and LRS). In Section \ref{sec3}, we investigate the physical nature of U13151 through a detailed analysis of its photometric and spectroscopic properties. In Section \ref{sec4}, we discuss our findings, from the rapid assembly of stellar mass through compact clumps to possible evolutionary pathways, including the connection between U13151-like systems and the progenitors of Blue Monsters at higher redshifts, as well as the early building blocks of present-day massive galaxies. Finally, we explore the potential contribution of U13151-like systems to Cosmic Reionization. In Section \ref{sec5}, we summarize our main findings and conclusions.

\vspace{2mm}
Throughout this paper, we consider a flat $\Lambda$CDM cosmology with $H_{0} = 67.7\; \rm km\;s^{-1}\;Mpc^{-1}$, $\Omega_{M} = 0.310$ (\citealt{collaboration_planck_2020}). All magnitudes are total and refer to the AB system \citep{oke_secondary_1983}. A \citet{chabrier_galactic_2003} initial mass function (IMF) is assumed. Moreover, in Table \ref{tab:emission_ratios}, we report the line ratios that will be adopted throughout this work. The magnification factor adopted for this source throughout the paper is $\mu=12.8\pm0.7$ (\citealt{furtak_uncovering_2023, price_uncover_2024}).

\begin{table}[h]
    \centering
    \renewcommand{\arraystretch}{1.5} 
    \begin{tabular}{l c}
        \hline
        Index & Definition \\
        \hline
        $\mathrm{R}3$ & $\log_{10} \left( \frac{ [\textsc{O iii}]\lambda5007 } {\textsc{H}\beta} \right)$ \\
        $\mathrm{O}32$ & $\log_{10} \left( \frac{ [\textsc{O iii}]\lambda\lambda 4959,5007 } { [\textsc{O ii}]\lambda\lambda 3727,3729} \right)$ \\
        $\mathrm{R}23$ & $\log_{10} \left( \frac{ [\textsc{O iii}]\lambda\lambda 4959,5007 + [\textsc{O ii}]\lambda\lambda 3727,3729}{\text{H}\beta} \right)$ \\
        $\mathrm{Ne}3\mathrm{O}2$ & $\log_{10} \left( \frac{ [\mathrm{Ne\,\textsc{iii}}]\lambda 3869 } { [\textsc{O ii}]\lambda\lambda 3727,3729} \right)$ \\
        $\mathrm{Ne}3\mathrm{O}2\mathrm{Hd}$ & $\log_{10} \left( \frac{ [\mathrm{Ne\,\textsc{iii}}]\lambda 3869 + [\textsc{O ii}]\lambda\lambda 3727,3729}{\text{H}\delta} \right)$ \\
        $\mathrm{O3Hg}$ & $\log_{10} \left( \frac{ [\textsc{O iii}]\lambda4363 } {\textsc{H}\gamma} \right)$ \\
        $\mathrm{O33}$ & $\log_{10} \left( \frac{ [\textsc{O iii}]\lambda5007 } { [\textsc{O iii}]\lambda4363 } \right)$ \\
        \hline
    \end{tabular}
    \caption{Definitions of line ratios adopted in this work. We use the O32 definition from \citet{calabro_evidence_2024}, that has a fixed offset of log$_{10}(1.3)$ between our values and the more common definition of O32 ($\equiv$ [O\,{\sc iii}]$\lambda$5007/[O\,{\sc ii}]$\lambda\lambda$ 3727,3729).}
    \label{tab:emission_ratios}
\end{table}

\begin{figure*}
    \centering
    \includegraphics[width=1.\linewidth]{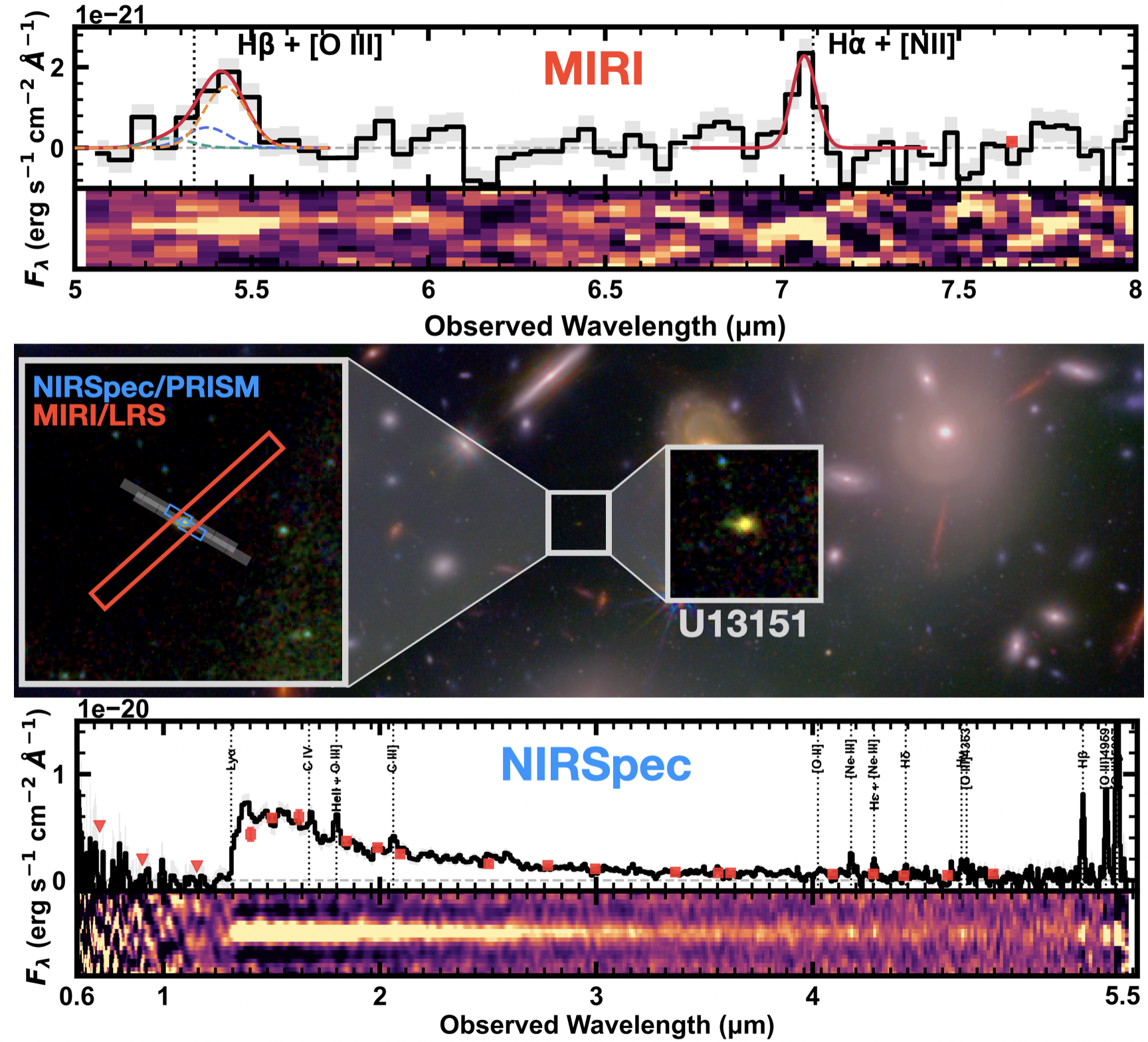}
    \caption{Overview of the spectroscopic data available for U13151, located behind Abell 2744. RGB are produced using F444W, F277W, and F115W (cutouts are: $5$\arcsec$\times\, 5$\arcsec\; and $1.5$\arcsec$\times\, 1.5$\arcsec). The {\bf upper panel} shows the MIRI/LRS spectrum from PID 8051, including both the 1D extraction and 2D spectra. The spectrum shows the H$\beta$+[O\,{\sc iii}] and H$\alpha$+[N\,{\sc ii}] emission-line complexes. Note that the vertical lines refer to the $z_{\rm spec}$ (9.7995) inferred from NIRSpec/PRISM (see Section 3.2.1). The {\bf lower panel} shows the NIRSpec/PRISM spectrum from PID 2561, again including both the 1D and 2D spectra. Several emission lines are highlighted, including H$\beta$ and [O\,{\sc iii}], which are fully resolved at the PRISM resolution. In the NIRSpec spectrum, the corresponding NIRCam photometry is shown as red squares (upper limits as upside triangles). Similarly, we show the MIRI photometry (only F770W) for the MIRI/LRS spectrum.}
    \label{fig:general_plot}
\end{figure*}

\section{Dataset}\label{sec2}

\subsection{MIRI}
\subsubsection{Imaging}

The MIRI imaging data (F770W and F1000W) are part of the {\it MIRI in Clusters} (MIRIC) program (PID 5578; PIs: E. Iani \& P. Rinaldi) and were reduced following the custom reduction procedure described in \citet{rinaldi_deciphering_2025}. Briefly, we start from the calibrated {\tt *\_cal.fits} products generated with the default {\tt jwst} pipeline\footnote{\url{https://zenodo.org/records/16280965}} ($\tt v1.20.0$, {\tt pmap: 1439}) with a set of fine tuned parameters in {\tt stage1}, after which custom extra-steps are used to improve the background modeling. An initial mosaic is first constructed and used to generate source masks using {\sc Source Extractor} (\citealt{bertin_sextractor_1996}), which are then reprojected onto the individual exposures to build a robust master background. This background is subtracted from each exposure after determining the optimal scaling factor, and residual row- and column-dependent patterns are further removed while masking detected sources. The procedure is iterated using refined source masks, including both compact and low-surface-brightness emission with {\sc NoiseChisel} (\citealt{akhlaghi_noise-based_2017}), before aligning the individual exposures to the UNCOVER astrometric reference frame and producing the final mosaics through outlier rejection and resampling steps. We refer the reader to \citet{rinaldi_deciphering_2025} for a detailed description of the MIRI data reduction, while the full presentation of the survey will be provided in a forthcoming paper by E. Iani et al. (in prep.). A similar procedure was adopted with other MIRI datasets (e.g., \citealt{alberts_smiles_2024, perez-gonzalez_what_2024, ostlin_miri_2024}). 

For F770W, we obtain a global $5\sigma$ depth of 27.10 mag within a $0.3$\arcsec-radius aperture, with the deepest region, overlapping the core of Abell 2744, reaching 27.43 mag. For F1000W, we obtain a global $5\sigma$ depth of 26.63 mag, with the deepest region reaching 26.90 mag\footnote{For the same integration times and aperture sizes, the ETC predicts $5\sigma$ depths of 26.40 mag in F770W and 26.23 mag in F1000W. Such improvements in depth with respect to the predictions from the ETC are consistent with those previously reported for ultra-deep MIRI imaging (see \citealt{ostlin_miri_2024}). For F770W, we adopt the empirical aperture correction adopted in \citet{robertson_jwst_2026} (see their Section 5.4). For F1000W, instead, we adopt an aperture correction derived from the encircled-energy files used by the pipeline ($r=0.3$\arcsec\,$\rightarrow1.72$).}.

\subsubsection{Low Resolution Spectroscopy}
The MIRI/LRS observations are part of {\it The PRImordial galaxy Survey with MIRI Spectroscopy} (PRISMS) program (PID 8051; PIs: J. \'{A}lvarez-M\'{a}rquez \& L. Colina) and were reduced following the procedure described in \citet{alvarez-marquez_prisms_2026}. Briefly, the data were processed with the JWST calibration pipeline ({\tt v1.20.2,
pmap: 1464}), using the standard Stage 1 and Stage 2 steps, with the cosmic-ray shower correction activated during the jump detection step. Before the final combination, several custom corrections were applied to the calibrated detector images, including wavelength masking, construction and subtraction of a master background, removal of residual wavelength-dependent background structure along the slit, and sigma clipping to reject remaining cosmic rays and detector artifacts. The final 2D spectrum and 1D extraction were then produced with Stage 3 of the pipeline, adopting an extraction aperture of 0.44 arcsec and applying the standard MIRI/LRS aperture correction under the assumption of a point-like source. We refer the reader to \citet{alvarez-marquez_prisms_2026} for a full description of the MIRI/LRS data reduction and calibration.

\subsection{Ancillary data}
\subsubsection{NIRCam}
In this work, we make use of the public UNCOVER imaging data release for Abell 2744 (\citealt{bezanson_jwst_2024}), which combines deep JWST/NIRCam observations from different programs. This includes JWST/NIRCam imaging from PID 2756 (PI: W. Chen), PID 2561, the {\it Ultra-deep NIRSpec and NIRCam Observations before the Epoch of Reionization} survey (UNCOVER; PIs: I. Labbé and R. Bezanson; \citealt{bezanson_jwst_2024}), PID 2883, MAGNIF (PI: F. Sun), PID 3538 (PI: E. Iani), PID 4111, the {\it Medium Bands, Mega Science} survey (MegaScience; PI: K. Suess; \citealt{suess_medium_2024}), and PID 3516, the {\it All the Little Things} program (ALT; PIs: R. Naidu and J. Matthee; \citealt{naidu_all_2024}). This extensive imaging coverage (with the following bands: F070W, F090W, F115W, F140M, F150W, F162M, F182M, F200W, F210M, F250M, F277W, F300M, F335M, F356W, F360M, F410M, F430M, F444W, F460M, F480M), makes Abell 2744 one of the deepest and best-characterized public JWST lensing fields to date. We refer the reader to \citet{suess_medium_2024} for more details on the dataset.

\subsubsection{NIRSpec}

The initial identification and spectroscopic confirmation of U13151 were based on the UNCOVER observations of Abell 2744 presented by \citet{fujimoto_uncover_2024}. The source was selected from the UNCOVER NIRCam imaging obtained in 2022 November (\citealt{bezanson_jwst_2024}), with photometry from the catalog of \citet{weaver_uncover_2024}. U13151 was subsequently targeted with JWST/NIRSpec MOS observations as part of the UNCOVER spectroscopic follow-up. We refer the reader to \citet{fujimoto_uncover_2024} for the full description of the target selection and observing strategy.

For the NIRSpec/MSA data presented in this paper, we have taken advantage of the most recent realization of the data reduction pipeline provided by the NIRSpec GTO team (described in \citealt{scholtz_jades_2025}), which extends the extracted and calibrated spectral traces on the detector significantly beyond the nominal red-end boundary (i.e., beyond $5.3\,\mu$m up to $5.5\,\mu$m) for each disperser/filter configuration. Compared to the initial reduction presented in \citet{fujimoto_uncover_2024}, this improved calibration allows the usable spectral coverage to extend far enough to include H$\beta$ and [O\,{\sc iii}]$\lambda\lambda4959,5007$.

\subsection{Absolute cross-instrument
flux calibration}

To account for residual uncertainties in the absolute spectrophotometric calibration and slit path-loss correction of the NIRSpec spectrum, we normalize the NIRSpec spectrum to the NIRCam photometry. 

For the photometry, we follow the approach outlined in \citet{rinaldi_deciphering_2025}. Briefly, we adopt circular apertures, where we adopted filter-dependent radii of $0.15$--$0.25$\arcsec, chosen to enclose the source emission while minimizing contamination from the intracluster light (ICL). We measured the local background in a $3$\arcsec$\times3$\arcsec\;region around the source and estimated the noise from non-adjacent pixels separated by five pixels to account for correlated noise, following \citet{perez-gonzalez_ceers_2023}. We corrected the measured fluxes to total using the aperture corrections provided by \citet{robertson_jwst_2026}.

We generate synthetic NIRSpec photometry by integrating the observed spectrum through the corresponding NIRCam broad- and medium-band transmission curves (using {\tt sedpy}\footnote{\url{https://github.com/bd-j/sedpy}.}) and compare these measurements with the observed photometric fluxes. A single multiplicative correction yields a normalization factor of $1.706\pm0.032$, such that $F_{\rm NIRSpec,corr}=1.706\,F_{\rm NIRSpec}$. 

We also tested whether a wavelength-dependent correction is required by fitting the ratio between the observed and synthetic NIRCam photometry with a first-order polynomial. To do so, we follow the approach outlined in \citealt{rinaldi_deciphering_2025}, building on the methodology outlined in \citealt{carniani_shining_2024}. The resulting correction is nearly flat, and differs from the single-factor normalization by $\lesssim1\%$ at H$\beta$ and [O\,{\sc iii}]$\lambda4959,5007$. We also tested the impact on the rest-frame UV continuum slope, finding a change of only $\lesssim1\%$ between the single-factor and wavelength-dependent normalizations.  We adopt the simpler single multiplicative normalization throughout this work; a similar result was obtained also by \citet{napolitano_gradual_2026} (see their Section 2.1).

Finally, we check the absolute flux calibration of the MIRI/LRS spectrum against the NIRSpec spectrum, after the latter has been calibrated to the NIRCam photometry. To do so, we independently fit the H$\beta$+[O\,{\sc iii}] emission-line complex in the NIRSpec/PRISM and MIRI/LRS spectra and compare the resulting integrated line fluxes, which agree within $1\sigma$. We therefore apply no additional correction to the MIRI/LRS spectrum. This is in line with other results presented in \citet{alvarez-marquez_prisms_2026}, \citet{marques-chaves_prisms_2026}, and \citet{perez-diaz_prisms_2026}. 

\section{On the nature of an ultra-faint galaxy in the early Universe}\label{sec3}
\begin{figure*}
\centering
\includegraphics[width=\textwidth]{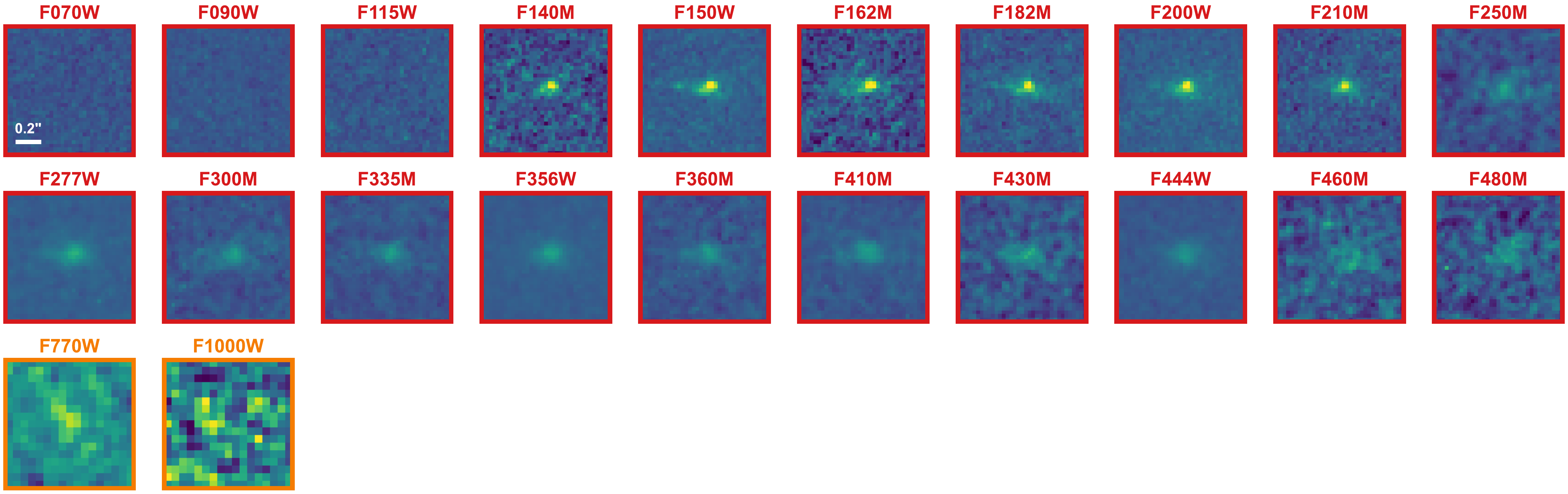}
\caption{NIRCam postage stamps of U13151 from F070W to F480M, together with MIRI postage stamps in F770W and F1000W. The NIRCam images are shown at 30 mas pixel scale, while the MIRI images are shown at 60 mas pixel scale. U13151 is not detected in F1000W, while it is detected in F770W due to the contribution of H$\alpha$ falling within this filter. Each stamp is 1\arcsec$\times$1\arcsec.}
\label{fig:imaging_overview}
\end{figure*}

In this section, we delve into both the photometric and spectroscopic properties of U13151. We present a spectroscopic overview of U13151 in Figure \ref{fig:general_plot}, together with an imaging overview in Figure \ref{fig:imaging_overview}.

\subsection{Insights from photometry and SED fitting}

We summarize all the photometric properties discussed in this section in Table \ref{tab:sed}.

\begin{deluxetable}{lcc}
\tablecaption{Photometry and SED-fitting Properties}
\tabletypesize{\fontsize{8.5}{10}\selectfont}
\tablewidth{0pt}
\tablehead{
\colhead{} & \colhead{Value} & \colhead{Error}
}
\startdata
\multicolumn{3}{c}{Photometry [nJy]} \\
\hline
F070W   & $<8$   & -- \\
F090W   & $<5$   & -- \\
F115W   & $<6$   & -- \\
F140M   & 28  & 4 \\
F150W   & 44  & 3 \\
F162M   & 52  & 6 \\
F182M   & 42  & 4 \\
F200W   & 41  & 3 \\
F210M   & 36  & 4 \\
F250M   & 33  & 5 \\
F277W   & 36  & 2 \\
F300M   & 32  & 4 \\
F335M   & 29  & 3 \\
F356W   & 29  & 1 \\
F360M   & 30  & 3 \\
F410M   & 33  & 3 \\
F430M   & 37  & 6 \\
F444W   & 29  & 2 \\
F460M   & 33  & 8 \\
F480M   & 47  & 7 \\
\hline
F770W   & 30  & 6 \\
F1000W  & $<35$  & -- \\
\hline
\multicolumn{3}{c}{SED-fitting properties} \\
\hline
$A_V$ [mag] & 0.14 & 0.10 \\
$\log_{10}(\mathcal{U})$ & $-1.57$ & 0.35 \\
Metallicity [$Z_{\odot}$] & 3.2\% & 1.5\% \\
$D_{4000}$ & 0.59 & 0.03 \\
$\mathrm{EW}_{0}(\mathrm{H}\alpha)$ [\AA] & 2333 & 798 \\
$\mathrm{EW}_{0}$([O\,{\sc iii}]$\lambda5007)$ [\AA] & 1354 & 183 \\
$\beta$ & $-2.55$ & 0.06 \\
Stellar Age$_{\rm young}$ [Myr] & 5.5 & 2.8 \\
SFR$_{10\,\mathrm{Myr}}$ [$M_{\odot}\,\mathrm{yr}^{-1}$]$^{*}$ & 0.49 & 0.13 \\
$\log_{10}(M_{\bigstar,\rm young}/M_{\odot})$$^{*}$ & 6.68 & 0.11 \\
$\log_{10}(M_{\bigstar,\rm old}/M_{\odot})$$^{*}$ & $<7.21$ & -- \\
$F$(H$\alpha$) & 170 & 40 \\
$F$([O\,{\sc iii}]$\lambda5007$) & 230 & 20 \\
\hline
\multicolumn{3}{c}{Morphology} \\
\hline
$R_{\rm e, F150W}$$^{*}$ [pc] & 122.4 & $^{+7.2}_{-7.9}$\\
\enddata
\tablecomments{
A “--” indicates that we report a $3\sigma$ upper limit.
SED-fitting properties are derived with \textsc{CIGALE}.
Emission-line fluxes are given in units of $10^{-20}\,\mathrm{erg\,s^{-1}\,cm^{-2}}$. $^{*}$ refers to quantities corrected for lensing magnification ($\mu = 12.8\pm0.7)$.
}
\label{tab:sed}
\end{deluxetable}

\subsubsection{The clumpy nature of U13151}

U13151 appears clumpy when observed in the short-wavelength channel from NIRCam. For this reason, we model it using {\sc lenstronomy} \citep{birrer_lenstronomy_2018}. We focus on the F150W imaging, which provides the best compromise among the short-wavelength filters in terms of angular resolution, depth, and SNR. At the redshift of U13151 ($z\approx9.8$), F150W probes the rest-frame UV emission of the system ($\approx0.14\,\mu$m).

\begin{figure}
    \centering
    \includegraphics[width=0.5\linewidth]{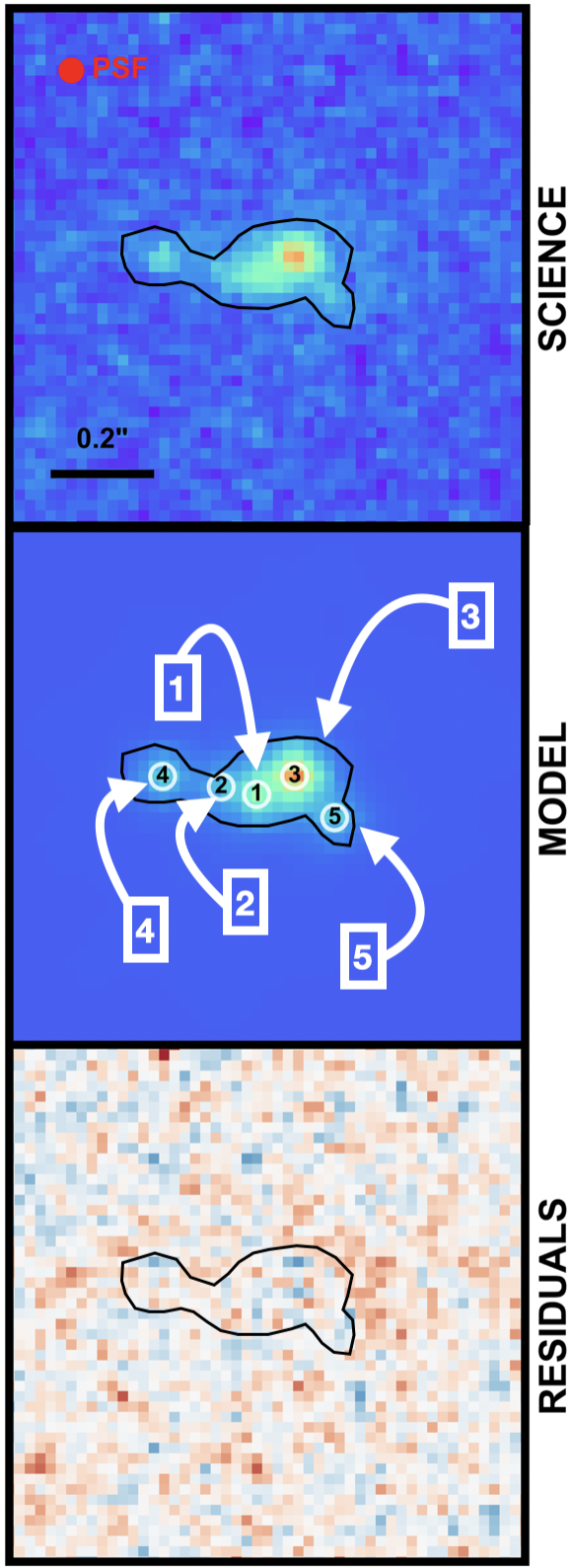}
    \caption{The F150W image (1\arcsec$\times$1\arcsec) used for the morphological decomposition. From top to bottom, the panels show the F150W science image, best-fitting model, and residuals. In the model panel, we mark the positions of the five compact clumps identified in the fit. We show in red the empirical FWHM of the NIRCam F150W filter.}
    \label{fig:f150w_decomposition}
\end{figure}

Briefly, we fit the observed F150W morphology with multi-component models designed to reproduce the compact UV-bright substructures in the galaxy; during the fit, the centroid of each component is left free to vary. We explored two alternative parameterizations, including one in which all components are modeled as point-like sources, and another in which the two brightest, partially extended components are modeled with Sérsic profiles. Among the models explored, the one invoking only point-like sources (5 in total) provides the best overall description of the data, yielding smaller residuals and a lower $\chi^2$. In particular, introducing Sérsic profiles results in more pronounced residuals around the brightest clump, without significantly improving the fit. More importantly, a comparison based on the Bayesian Information Criterion (BIC; \citealt{liddle_information_2007}) shows that the simpler, all-point-source model is statistically preferred ($\Delta_{\rm BIC}\gtrsim100$), indicating that the additional free parameters introduced by the Sérsic profiles are not justified by the data. The best-fitting model is shown in Figure~\ref{fig:f150w_decomposition}. 

Our results are consistent with the clumpy rest-frame UV structure often observed in star-forming galaxies at high redshift, where compact UV-bright knots trace localized regions of recent star formation (e.g., \citealt{guo_clumpy_2015, claeyssens_star_2023}). Nonetheless, we stress that the current data might only reveal the highest-surface-brightness UV components. A fainter, more diffuse underlying stellar component could still be present below the current surface-brightness limit (see discussion in \citealt{rinaldi_beyond_2025}). In principle dust can be another reason behind why U13151 could look clumpy; though, as we will discuss later, this system is dust free.

Finally, we estimate the F150W effective radius of the full system. This radius encloses 50\% of the total F150W light and therefore traces the spatial extent of the observed rest-frame UV emission of U13151. We find an observed $R_{\rm e}=0.107_{-0.004}^{+0.003}$\arcsec, corresponding to $R_{\rm e}=450.4_{-13.7}^{+15.5}$ pc at $z=9.7995$. Assuming isotropic magnification ($\mu=12.8\pm0.7$), this gives an intrinsic (corrected for PSF as well) effective radius of $R_{\rm e}=122.4_{-7.9}^{+7.2}$ pc. This size is consistent with the recent analysis of \citet{ono_morphological_2025}, who found $R_{\rm e}\approx30\text{--}500$ pc for galaxies at $z>10$, with a median size of $133_{-12}^{+13}$ pc. We will discuss this further in Section 4.1. We remind the reader that the F150W PSF has a FWHM of $\approx0.05$\arcsec, corresponding to a source-plane resolution of $\approx30$ pc after correcting for lensing (i.e., considering its HWHM); this means that the individual clumps within U13151 are unresolved below $30$~pc (comparable to compact star-forming structures seen in other strongly lensed high-redshift galaxies; e.g., \citealt{claeyssens_star_2023, claeyssens_tracing_2025}).

\subsubsection{Spectrophotometric view of U13151}

We now derive the stellar properties of U13151 through SED fitting using both photometry and spectroscopy. To assess the physical properties of U13151, we modeled the SED with the spectrophotometric version of \textsc{CIGALE} (\citealt{boquien_cigale_2019}). We adopted the general setup of \citet{alvarez-marquez_prisms_2026}, using the Bruzual \& Charlot stellar population models (\citealt{bruzual_stellar_2003}) and a Chabrier IMF (\citealt{chabrier_galactic_2003}). Nebular continuum and line emission were included assuming $n_{\rm e}=1000~{\rm cm^{-3}}$, $-3.2\leq\log U\leq-1.3$, $f_{\rm esc}=0$, and stellar and nebular metallicities spanning $0.02$--$0.2\,Z_\odot$\footnote{We adopt an electron density of $n_e=1000\,\mathrm{cm}^{-3}$, as the CIGALE nebular models do not allow us to adopt $n_e=300\,\mathrm{cm}^{-3}$ which we use in Section 3.2. We do not expect this assumption to significantly affect the inferred stellar population properties, although some dependence of the nebular emission-line ratios on electron density cannot be excluded.}. These parameter ranges were constrained, where possible, by the physical conditions inferred from the observed spectral properties of U13151 (see Section 3.2). Dust attenuation is modeled using the {\sc CIGALE} module {\tt dustatt\_modified\_starburst}. Stellar attenuation follows the modified Calzetti prescription of \citet{noll_analysis_2009}, based on the Calzetti law \citep{calzetti_dust_2000}. We vary the attenuation curve slope $\delta$ between $-0.6$ and $0.4$, and fix the UV bump amplitude to zero. Nebular dust attenuation is modeled using the Small Magellanic Cloud (SMC) extinction curve of \citet{pei_interstellar_1992}. The nebular color excess $E(B-V)_{\rm neb}$ ranges from 0 to 0.3 mag, as inferred from the Balmer decrement (see Section 3.2.3), which is consistent with negligible attenuation but allows values as high as $\approx0.3$ mag within the uncertainties. The stellar-to-nebular dust attenuation ratio ranges from 0.33 to 1. For the star formation history (SFH), we adopted two constant star-forming components to account for both a young stellar population and a potentially more evolved underlying population. The age of the young component is allowed to vary between 1 and 10 Myr, while that of the older component spans 10--200 Myr. This parametrization allows us to investigate the contribution of an underlying older stellar population while independently constraining the recent star-forming activity traced by the strong nebular emission.

From \textsc{CIGALE}, we infer a very low stellar mass for the young stellar component of $\log_{10}(M_{\bigstar}/M_{\odot}) = 6.68\pm0.11$ (after correcting for magnification), placing U13151 among the least massive young stellar systems currently observed at $z\gtrsim10$. Although our SED modeling also allows for an underlying older stellar population, its contribution remains largely unconstrained by the available data. In particular, the non-detection of U13151 in MIRI/F1000W limits our ability to constrain the rest-frame optical continuum and, consequently, to establish the presence and stellar mass of a potentially older population (see Figure \ref{fig:imaging_overview}). We therefore place a $3\sigma$ upper limit of $M_{\bigstar\,\rm old}<1.64\times10^7\,M_\odot$ on this component and, throughout the following analysis and discussion, adopt the stellar mass inferred for the young component as our reference value. The SED fitting further suggests negligible dust attenuation, with $A_V=0.14\pm0.10$ mag, consistent with virtually no dust content within the uncertainties (and in agreement with the observed Balmer decrement; see next Section). Moreover, we infer a remarkably young stellar age of $5.5\pm2.8$ Myr, suggesting that U13151 may be undergoing its first major episode of star formation. 
The inferred instantaneous SFR (${\rm SFR}_{10,{\rm Myr}}$) is $0.49\pm0.13~M_{\odot}\,{\rm yr}^{-1}$ (after correcting for magnification), corresponding to a specific SFR (sSFR) of $103\pm37\,{\rm Gyr}^{-1}$. The inferred metallicity is also remarkably low, with $Z=(3.2\pm1.5)\%\,Z_{\odot}$. We also obtain a UV slope of $\beta=-2.55\pm0.06$. Finally, we retrieve $\log_{10}(\mathcal{U})=-1.57\pm0.35$, highlighting the extreme ionization conditions of the system.

\begin{figure*}
    \centering
    \includegraphics[width=1.\linewidth]{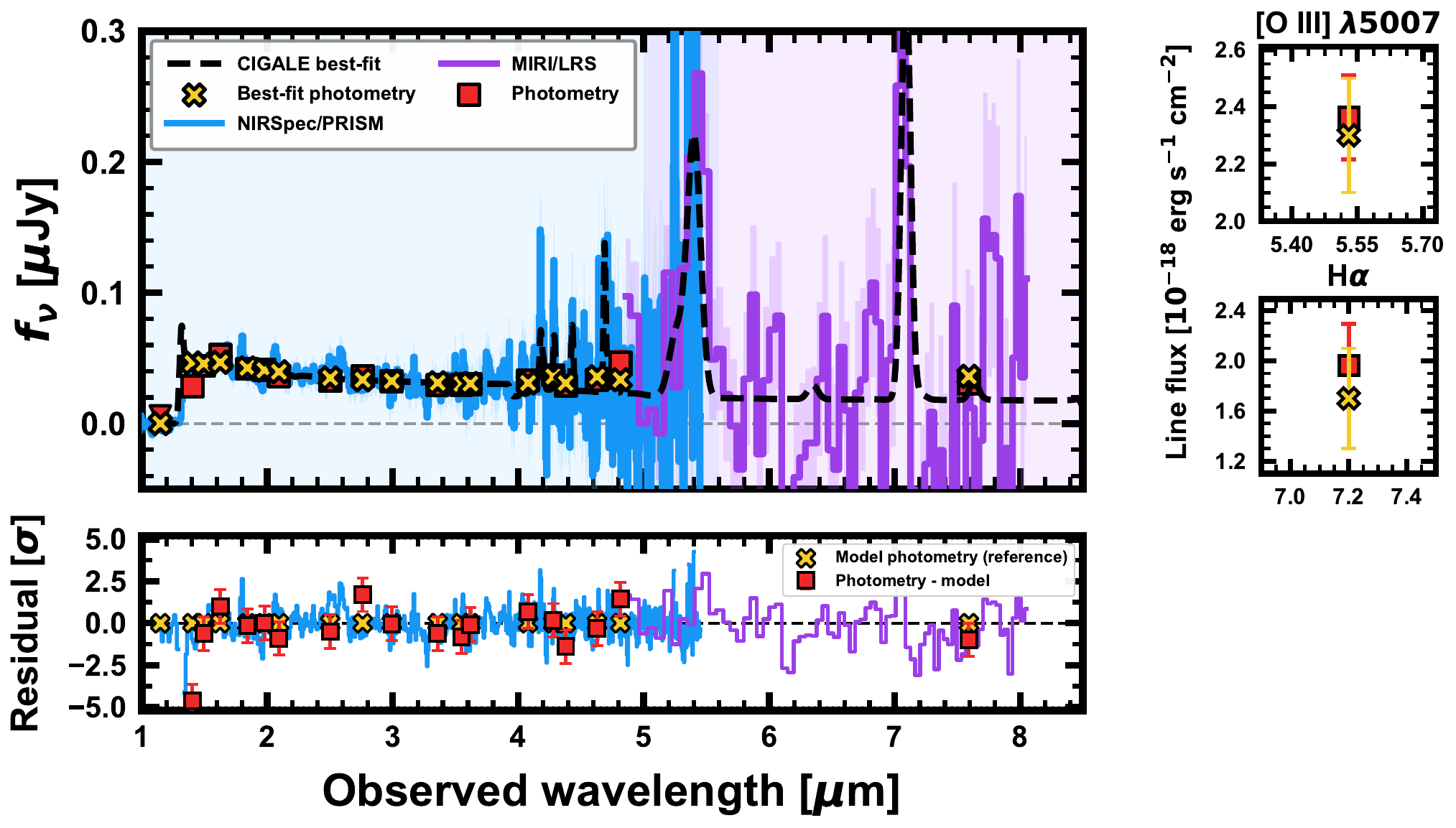}
    \caption{Best-fit SED fit of U13151. The SED fit analysis combine the NIRCam photometry, NIRSpec/PRISM spectrum along with [O\,{\sc iii}]$\lambda5007$ flux, and MIRI photometry along with MIRI/LRS derived H$\alpha$ flux. Upper limits ($3\sigma$) are shown as triangles. The best-fit CIGALE model is represented with dashed black line, and its synthetic NIRCam and MIRI photometry, are represented by yellow crosses.}
    \label{fig:sedfitting}
\end{figure*}

We will show later in the text (Section 3.2) that the properties inferred from the SED fitting align closely with those derived directly and independently from NIRSpec/PRISM and MIRI/LRS. Indeed, as a sanity check, the emission-line fluxes predicted by \textsc{CIGALE} are consistent with those measured spectroscopically within $\approx1\sigma$, further supporting a scenario in which the observed emission is dominated by a very young, actively star-forming stellar population. We show the best-fit SED from {\sc CIGALE} in Figure \ref{fig:sedfitting} and summarize the SED-fitting (bayesian) properties in Table \ref{tab:sed}.

\subsection{Insights from spectroscopy}
We summarize all the spectroscopic properties discussed in this section in Table \ref{tab:spec}. All emission line measurements have been carried out using {\sc MSAEXP} \citep{brammer_msaexp_2023, heintz_jwst-primal_2025}\footnote{For MIRI/LRS, we provided {\sc MSAEXP} with the appropriate spectral resolution curve.}. At $z\gtrsim8$, rest-frame optical lines fall at $\gtrsim4\,\mu$m, where the PRISM resolving power is highest and can exceed nominal values for compact sources (see discussion in \citealt{de_graaff_ionised_2024}). H$\gamma$ and [O\,{\sc iii}]$\lambda4363$, as well as [Ne\,{\sc iii}]$\lambda3869$ and the adjacent H8+He\,{\sc i}$\lambda3889$ complex, can therefore be separately constrained through simultaneous profile fitting (\citealt{deugenio_jades_2024, hsiao_jwst_2024}).

\begin{deluxetable}{lcc}
\tablecaption{Emission Line Fluxes and Derived Spectral Properties}
\tabletypesize{\fontsize{8.5}{10}\selectfont}
\tablewidth{0pt}
\tablehead{
\colhead{Line} & \colhead{Value} & \colhead{Error}
}
\startdata
N\,{\sc iv}]$\lambda1486$ & $<67$ & \\
C\,{\sc iv}$\lambda$1549 & 60.9 & 20.2 \\
O\,{\sc iii}]$\lambda$1660,1666 + He\,{\sc ii}$\lambda1640$ & 78.5 & 17.2 \\
N\,{\sc iii}]$\lambda1750$ & $<46$ & \\
C\,{\sc iii}]$\lambda$1907,1909 & 41.9 & 13.1 \\
{[O\,{\sc ii}]}$\lambda\lambda$3727,3729 & 7.7 & 3.7 \\
{[Ne\,{\sc iii}]}$\lambda$3869 & 24.1 & 3.3 \\
He\,{\sc i}$\lambda3889$ + H8  &      9.6 &  2.6 \\
{[Ne\,{\sc iii}]}$\lambda$3968 + H$\epsilon$ & 15.5 & 2.8 \\
H$\delta$ & 11.4 & 5.5 \\
H$\gamma$ & 21.9 & 9.0 \\
{[O\,{\sc iii}]}$\lambda$4363 & 8.6 & 4.3 \\
H$\beta$ & 68.1 & 10.5 \\
{[O\,{\sc iii}]}$\lambda$4959 & 78.5 & 4.9 \\
{[O\,{\sc iii}]}$\lambda$5007 & 236.3 & 14.7 \\
\hline
H$\beta$ + [O\,{\sc iii}]$^{***}$ & 342.6 & 61.1 \\
H$\alpha$ + [N\,{\sc ii}]$^{***}$ & 196.1 & 33.3 \\
\hline
\hline
\hline
Galaxy Properties & & \\
\hline
\hline
$A_V$$^{\dagger}$ (H$\alpha$/H$\beta$) [mag] & $0.1$& $0.62$ \\
12+log(O/H)$_{\rm SL}$ & 7.29  & $_{-0.05}^{+0.05}$ \\
12+log(O/H)$_{\rm Te}$ & 7.44  & $_{-0.17}^{+0.33}$ \\
$\rm log_{10}(C/O)$ & $-0.75$  & 0.19 \\
$\rm log_{10}(Ne/O)$ & $-0.51$  & 0.07 \\
$T_{e}(O^{+})$ [K] &16288 &  $^{+2844}_{-3612}$\\ 
$T_{e}(O^{2+})$ [K] & 18983 &  $^{+4063}_{-5160}$\\
$\text{log}_{10}(\mathcal{U})^{**}$ & $-1.90$ & 0.35  \\
$\text{EW}_{0}$([O\,{\sc iii}]$\lambda\lambda4959,5007$) [\AA] & $>1045$  & --  \\
$\text{EW}_{0}(\text{H}\alpha)$ [\AA] & $>938$  &  -- \\
SFR(H$\alpha$) [$M_\odot\,\mathrm{yr}^{-1}$]$^{*}$ & 0.45 &  0.08 \\
SFR(UV) [$M_\odot\,\mathrm{yr}^{-1}$]$^{*}$ & 0.29 & 0.02 \\
$\mathcal{B}$ [SFR(H$\alpha$)/SFR(UV)] & 1.58 & 0.27  \\
$M_{UV}*$ [mag] & $-17.44$  & 0.07  \\
$\beta$ & $-2.64$  & 0.15 \\
$f_{esc,LyC}$ & 0.19  & 0.17  \\
$\xi_{ion}$ \, $[\log_{10}(\mathrm{Hz\, erg}^{-1})]$ & 25.64  & 0.12  \\
\hline
$z_{spec}$ & 9.7995  & 0.0270  \\
\enddata
\tablecomments{Observed fluxes and errors (not corrected for lensing magnification, but we provide its value from \citealt{furtak_jwst_2023, fujimoto_uncover_2024, price_uncover_2024}; $\mu=12.8\pm0.7$) are given in units of $10^{-20} \, \mathrm{erg\,s^{-1}\,cm^{-2}}$. Derived galaxy properties with associated uncertainties are provided. Uncertainties were estimated from the posterior distributions sampled with the Python package {\sc emcee}, using 10000 MCMC steps. $^{*}$ refers to quantities corrected for lensing effects. $^{**}$ refers to the estimated median quantities. $^{***}$ The reported flux corresponds to the unresolved complexes in LRS: H$\beta$ + [O\,{\sc iii}] and H$\alpha$ + [N\,{\sc ii}]$\lambda\lambda6548,6583$. However, given the very low metallicity of the source, the expected [N\,{\sc ii}] contamination is negligible, at the level of $\approx1\text{--}2\%$ based on the estimates of \citet{anders_spectral_2003}. ${\dagger}$ we adopt $A_V=0$ in the rest of the analysis.}
\label{tab:spec}
\end{deluxetable}

\subsubsection{Redshift (re)assessment of U13151}
The NIRSpec/PRISM data for U13151 were first reported by \citet{fujimoto_uncover_2024} at $z_{\rm spec}=9.880_{-0.001}^{+0.012}$, based on early results. Updated NIRSpec calibrations extend the wavelength solution to longer wavelengths, enabling the recovery of [O\,{\sc iii}]$\lambda\lambda4959,5007$ at such high redshifts. Moreover, their spectrum already showed evidence for strong damping-wing absorption, which {\it could} bias redshift estimates when not anchored by strong emission lines, as evident in their Figure 3. We re-estimated the redshift with {\sc MSAEXP}. {\sc MSAEXP} fits the observed spectrum using a combination of Gaussian continuum splines and emission-line templates. We adopted {\tt nsplines = 11} and searched for the minimum reduced $\chi^{2}$ over the redshift interval $z = 9$--$10$, motivated by the previous spectroscopic estimate from \citet{fujimoto_uncover_2024}. We derive a lower redshift than previously reported, finding $z_{\rm spec} = 9.7995 \pm 0.0270$\footnote{The quoted uncertainty includes in quadrature the statistical fitting uncertainty and the nominal NIRSpec/MSA PRISM wavelength-calibration uncertainty of $\approx750\,{\rm km\,s^{-1}}$, corresponding to $\sigma_z\approx0.027$ at $z\approx9.8$ (\citealt{boker_-orbit_2023}).}.

Finally, we also derived $z_{\rm spec}$ independently from the MIRI/LRS spectrum. In particular, from the H$\beta$ + [O\,{\sc iii}]$\lambda\lambda4959,5007$ complex, we obtain $z_{\rm spec}=9.8351_{-0.0436}^{+0.0413}$, while H$\alpha$ yields $z_{\rm spec}=9.7621_{-0.0314}^{+0.0318}$. At the spectral resolution of MIRI/LRS, the H$\beta$+[O\,{\sc iii}] and H$\alpha$+[N\,{\sc ii}] complexes are unresolved, introducing additional uncertainties in the inferred redshifts. Moreover, the wavelength calibration of MIRI/LRS is subject to systematic uncertainties of up to $\approx20$ nm, as reported by STScI\footnote{See \url{https://jwst-docs.stsci.edu/jwst-calibration-status/miri-calibration-status/miri-lrs-calibration-status}.}. These limitations may account for the small offsets between the redshifts inferred from MIRI/LRS and NIRSpec/PRISM, which nonetheless remain consistent within $1\sigma$. Throughout this work, we adopt the $z_{\rm spec}$ derived from NIRSpec/PRISM.

\subsubsection{Presence of a strong damping-wing absorption}
As discussed above, the higher redshift reported by \citet{fujimoto_uncover_2024} may be driven by damping-wing absorption shifting the apparent Ly$\alpha$ break redward. We tested this by fitting the Ly$\alpha$ absorption following \citet{deugenio_jades_2024} and \citet{hainline_searching_2024}\footnote{Code available at: \url{https://github.com/joriswitstok/lymana_absorption}.}. During the fit, we fixed the systemic redshift to the value inferred from the emission lines. We also modeled the Ly$\alpha$ region by including IGM absorption and a damped Ly$\alpha$ absorber with free $N_{\rm HI}$, convolved with the instrumental line-spread function. We find a Ly$\alpha$-break redshift of $z=9.814^{+0.047}_{-0.044}$, slightly above the systemic value, as expected for damping-wing absorption. 

The absorber redshift, $z_{\rm DLA}=9.799^{+0.050}_{-0.035}$, is consistent with $z_{\rm sys}\approx9.7995$, suggesting gas associated with the galaxy or its circumgalactic medium (CGM) rather than an unrelated foreground absorber. The fit favors $\log(N_{\rm HI}/{\rm cm}^{-2})=21.35^{+0.34}_{-0.50}$, firmly above the classical DLA threshold \citep{wolfe_damped_2005}, though below the most extreme systems at similar redshifts \citep[e.g.,][]{heintz_strong_2024}. The fit allows values around $x_{\rm HI}\approx0.7$ and $R_{\rm ion}\approx9$ pMpc, although both remain highly degenerate with $N_{\rm HI}$, absorber redshift, the intrinsic continuum, and the adopted damping-wing prescription. We stress that these values as indicative only and do not interpret $R_{\rm ion}$ as a direct measurement of the physical size of an ionized bubble surrounding the source.

\begin{figure}
    \centering
    \includegraphics[width=1.\linewidth]{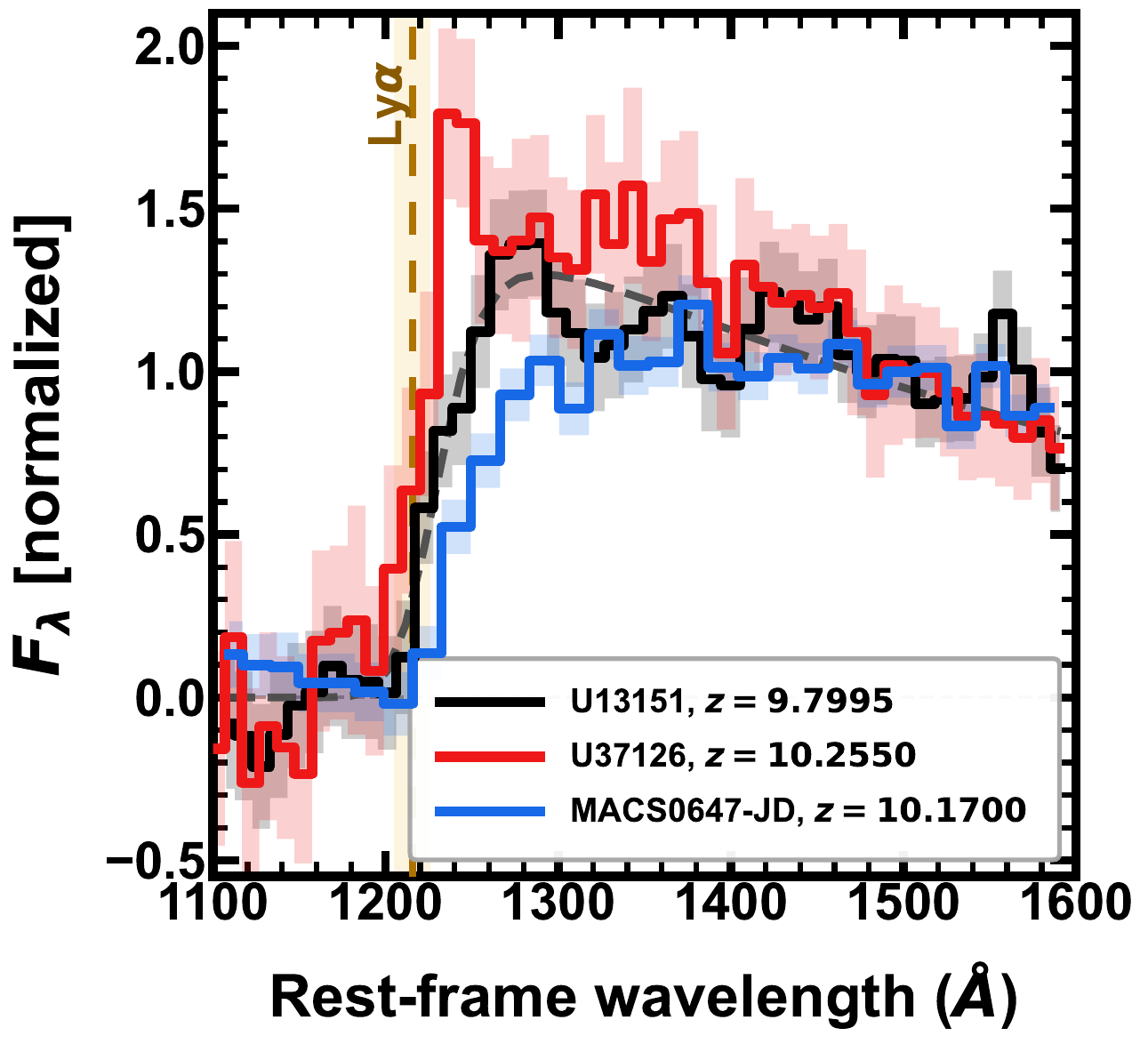}
    \caption{The spectrum of U13151 (black) compared to other high-$z$ galaxies, namely U37126 (\citealt{marques-chaves_prisms_2026}) and MACS0647-JD (\citealt{heintz_strong_2024}), with the latter showing a strong damped Ly$\alpha$ absorption feature ($\rm N_{HI}\approx2.5\times 10^{22}\; \rm cm^{-2}$). In the case of U13151, we find $\rm N_{\rm HI}=2.24^{+2.66}_{-1.53}\times10^{21}\ {\rm cm}^{-2}$ (model shown as a dashed black line). All spectra have been normalized at $\lambda \approx 1450\text{--}1550$\,\AA.}
    \label{fig:damping}
\end{figure}

Nonetheless, we interpret the fit mainly as evidence that the Ly$\alpha$ break is consistent with substantial neutral hydrogen absorption from the galaxy/CGM and possibly the surrounding IGM. Figure \ref{fig:damping} compares U13151 with U37126 \citep[red;][]{marques-chaves_prisms_2026} and MACS0647-JD \citep[blue;][]{heintz_strong_2024}, the latter having $\rm N_{HI}\approx2.5\times10^{22}\,cm^{-2}$. Recently, a similar (though more extreme) case of large column density was reported in GHZ2, where \citet{castellano_investigating_2026} linked it to the stringent Ly$\alpha$ non-detection (as in the case of U13151) and interpreted it as evidence for a large gas reservoir fueling compact, intense star formation during rapid galaxy assembly.

\subsubsection{Balmer Decrement and Dust Content}
Since U13151 is covered by both NIRSpec/PRISM and MIRI/LRS, we have access to several Balmer emission lines, namely H$\alpha$ from MIRI/LRS, and H$\beta$, H$\gamma$, and H$\delta$ from NIRSpec/PRISM. The observed Balmer decrement between H$\alpha$ and H$\beta$ is H$\alpha$/H$\beta = 2.88\pm0.64$. Because MIRI/LRS does not resolve H$\alpha$ from the [N\,{\sc ii}]$\lambda\lambda6548,6583$ doublet, we considered the correction from \citet{anders_spectral_2003} for extremely low-metallicity systems ($\lesssim5\%\;Z_{\odot}$; see next sections). In this metallicity regime, the expected [N\,{\sc ii}] contribution is very small, with the corrected H$\alpha$ flux corresponding to $\approx98\text{--}99\%$ of the observed H$\alpha$+[N\,{\sc ii}] (see also, e.g., \citealt{izotov_mmt_2007}). This correction is therefore negligible compared to the observational uncertainty, and we do not apply it in the following analysis. In any case, this would push toward a lower Balmer decrement. 

Assuming Case B recombination \citep{osterbrock_astrophysics_2006}, and adopting the electron temperature derived in the next section, we use \textsc{PyNeb} \citep{luridiana_pyneb_2015} to compute the intrinsic Balmer ratios. For our fiducial electron density\footnote{Following the assumption adopted in \citealt{isobe_jades_2026}, who use this value as a fiducial electron density when the [S\,{\sc ii}]$\lambda\lambda6716,6731$ doublet is not detected.} of $n_{\rm e}=300~{\rm cm^{-3}}$, the $T_{\rm e}(O^{2+})\approx1.9\times10^4$ K gives an intrinsic H$\alpha$/H$\beta\approx2.74$. The result is essentially insensitive to the assumed density over the range considered.  Assuming the Small Magellanic Cloud (SMC) reddening law \citep{gordon_quantitative_2003}, we find $A_{V}=0.1\pm0.62$~mag. In particular, given the very blue UV slope derived for U13151, we conclude there is no evidence for nebular dust attenuation and adopt $A_V=0$ mag throughout this work.

As a final check, we also examined the higher-order Balmer lines, H$\gamma$ and H$\delta$, for which the expected theoretical ratios are H$\gamma$/H$\beta=0.475$ and H$\delta$/H$\beta=0.264$ under Case B recombination \citep{osterbrock_astrophysics_2006}. However, given their relatively low SNR ($\approx2$--$3\sigma$), these measurements provide weaker constraints on dust attenuation compared to H$\alpha$ and H$\beta$. Therefore, we retain the attenuation inferred from the H$\alpha$/H$\beta$ Balmer decrement, which is consistent with negligible dust attenuation, and adopt $A_V=0$ mag throughout this work.

\subsubsection{UV-$\beta$ slope, Burstiness, and Ionizing Photon Production Efficiency}

\begin{figure*}
    \centering
    \includegraphics[width=\textwidth]{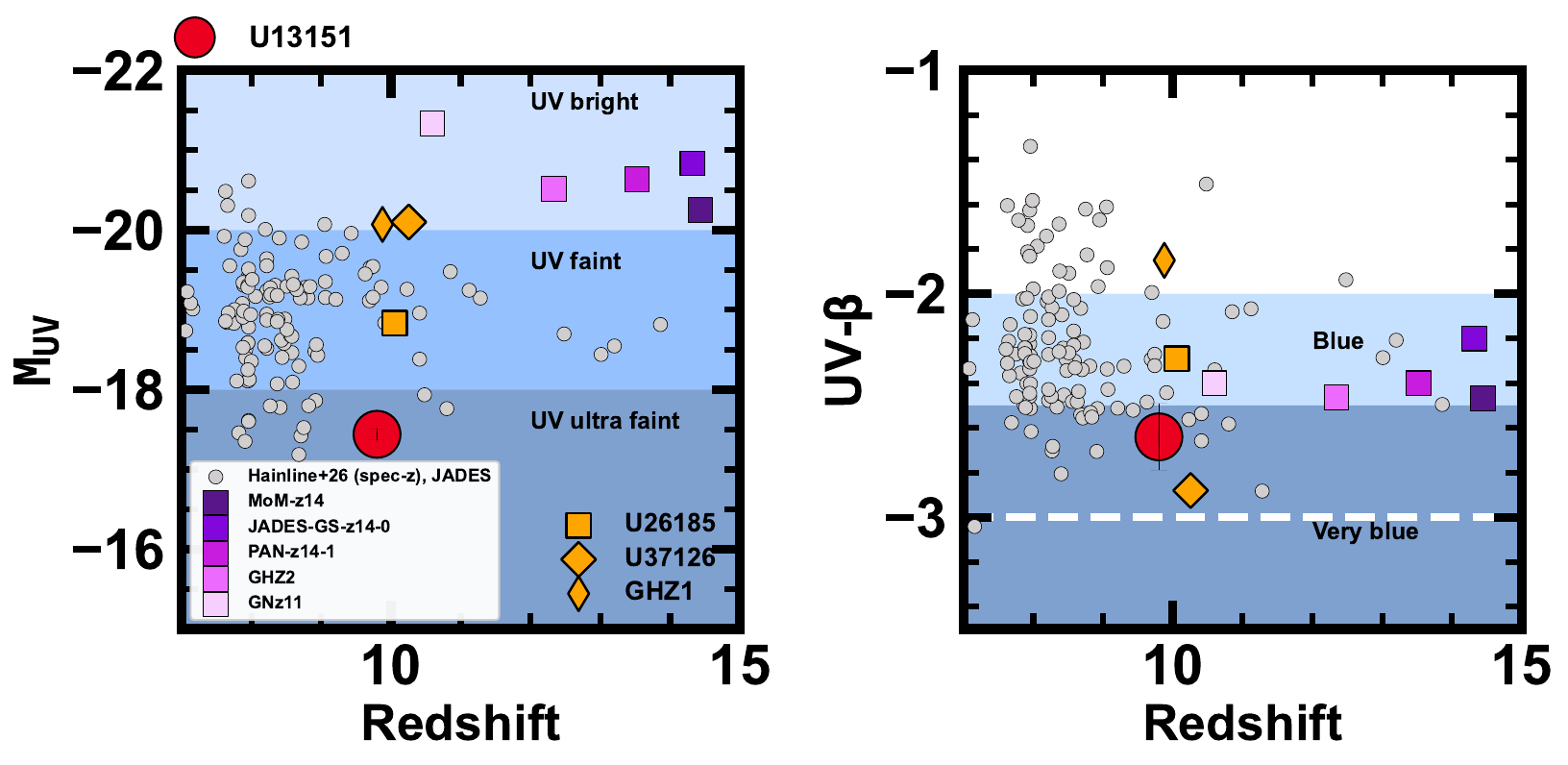}
    \caption{{\bf Left panel:} $M_{\rm UV}$ as a function of redshift. {\bf Right panel:} UV-$\beta$ as a function of redshift; we also show, as white dashed horizontal line, the maximum UV slope from stellar models. In both panels, U13151 is shown as a red filled circle. Gray points show the latest spectroscopically confirmed compilation in GOODS-North and GOODS-South presented by \citet{hainline_jwst_2026}. We also include several known UV-bright galaxies: GN-z11 (\citealt{bunker_jades_2023}), GHZ2 (\citealt{castellano_jwst_2024, zavala_luminous_2025}), PAN-z14-1 (\citealt{donnan_spectroscopic_2026}), JADES-GS-z14-0 (\citealt{carniani_shining_2024}), and MoM-z14 (\citealt{naidu_cosmic_2026}). U13151 is the first ultra-faint galaxy being observed at such high redshift with both MIRI imaging and LRS. We also show other galaxies from the PRISMS survey (U26185, U37126, and GHZ1; \citealt{alvarez-marquez_prisms_2026, marques-chaves_prisms_2026, perez-diaz_prisms_2026}).} 
    \label{fig:UV_prop}
\end{figure*}

One of the main motivations for following up U13151 with MIRI/LRS was its extreme faintness, together with its very blue UV-$\beta$. Based on previous measurements, \citet{fujimoto_uncover_2024} reported an intrinsic UV absolute magnitude of $M_{\rm UV}=-17.63$ for this object. We find a consistent but slightly fainter value of $M_{\rm UV}=-17.44\pm0.07$~mag (observed $M_{UV} = -20.21\pm0.1$~mag), with independent estimates from the spectrum and photometry agreeing within the uncertainties. This places U13151 in the sub-$L_{\bigstar}$ regime ($\approx20$ times fainter than $L_{\bigstar}$ at $z\approx10$; \citealt{harikane_comprehensive_2023}), and the faintest systems observed at $z\gtrsim10$ (see Figure \ref{fig:UV_prop}).

The source is also characterized by a very blue UV slope ($1350\text{--}2200$\AA) with $\beta_{\rm UV}=-2.64\pm0.15$ measured from the spectrum, in agreement with the value inferred from the photometry. We show U13151 along with the recent literature of high-$z$ galaxies ($z>8$) in Figure \ref{fig:UV_prop}. Such a blue UV-$\beta$ is compatible with no extinction and very young ages (e.g, \citealt{katz_megatron_2026}).

The combination of the NIRSpec/PRISM and MIRI/LRS spectra provides an unprecedented continuous rest-frame UV-to-optical coverage for such a faint system like U13151, allowing us to directly estimate the star-formation rate (SFR) from both H$\alpha$ and the UV continuum (at 1500\,\AA). To do so, we make use of the calibrations from \citet{kennicutt_star_2012} based on a Kroupa IMF\footnote{Note that the difference between a Kroupa and a Chabrier IMF is only about 0.03 dex, we therefore treat them equally here.}. Since U13151 is very metal poor (see Section 3.2.5), we use the metallicity-dependent corrections from \citet{theios_dust_2019}, adopting their lowest-metallicity calibration ($\approx 14\%$ solar; see their Table 2) for both UV and H$\alpha$. We note, however, that this metallicity is still approximately a factor of two higher than the value inferred for U13151. We also explored the correction factors provided by \citet{reddy_effects_2022} ($\approx 7\%$ solar) which provide consistent results, as also reported by \citet{alvarez-marquez_prisms_2026} for U26185.

\begin{figure}
    \centering
    \includegraphics[width=1.\linewidth]{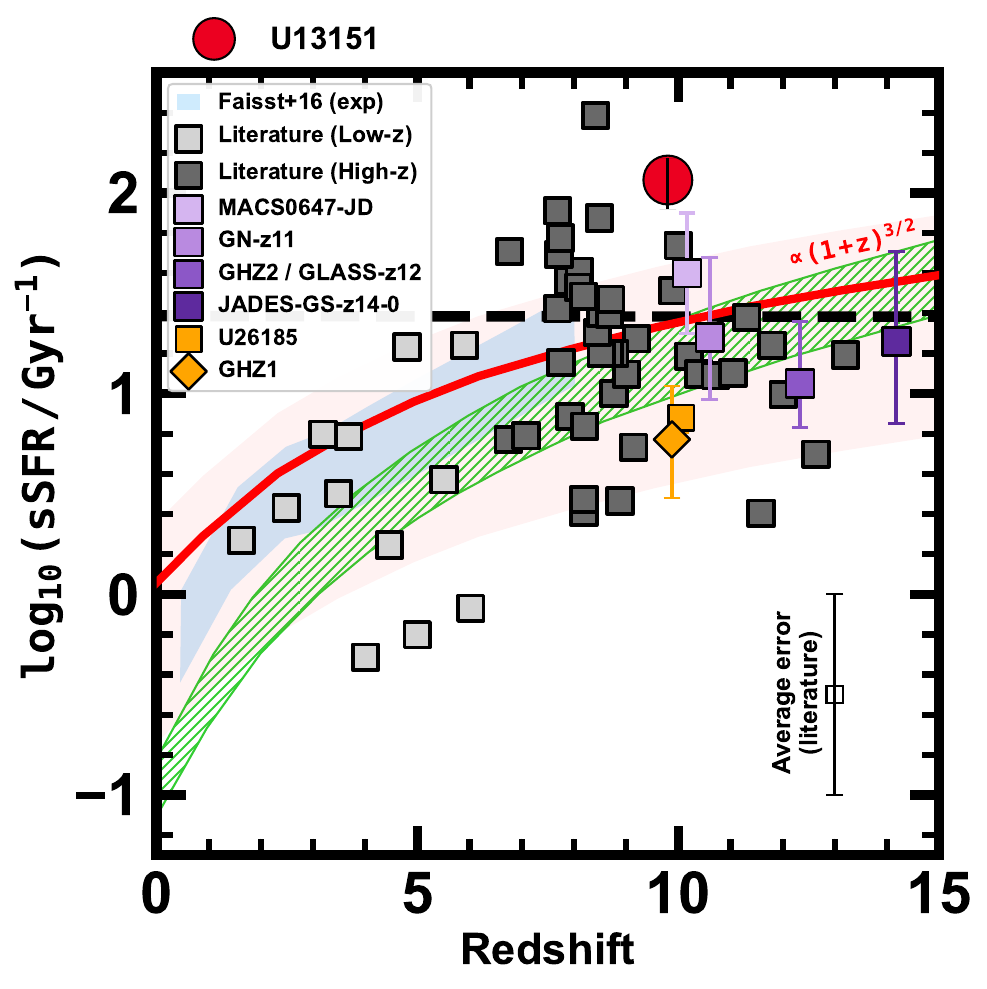}
    \caption{Redshift evolution of the sSFR. Literature at both low and high redshift is shown as light and dark gray squares (data points are from Figure 2 in \citealt{ferrara_super-early_2024}). We also show some of the galaxies with recent detection of H$\alpha$ at $z>10$ with MIRI (\citealt{hsiao_jwst_2023, alvarez-marquez_insight_2025, zavala_luminous_2025, helton_ionizing_2026}). In shaded blue, we show the evolution from \citet{faisst_coherent_2016}, while in green the expected sSFR evolution if stellar growth tracks the cosmological assembly of dark-matter halos, for systems that evolve into present-day halos with $\rm log_{10}(M_{0}/M_{\odot})\approx10\text{--}15$ (see \citealt{ferrara_super-early_2024}). Finally, in red the proposed evolution ($\propto(1+z)^{3/2}$) by \citet{ferrara_super-early_2024}. The black line refers to the critical sSFR derived by \citet{ferrara_super-early_2024}. U13151 is remarkably above the population of galaxies at $z>10$, which highlights its extreme nature. For comparison, we also show U26185 and GHZ1 from the PRISMS survey (\citealt{alvarez-marquez_prisms_2026, perez-diaz_prisms_2026}). We note that for GHZ1, we report the sSFR inferred from SED fitting.
}
    \label{fig:ssfr}
\end{figure}

From H$\alpha$, we infer ${\rm SFR}_{\rm H\alpha}=0.45\pm0.08~M_\odot\,{\rm yr}^{-1}$. Independently, the UV continuum (at 1500~\AA) gives ${\rm SFR}_{\rm UV}=0.29\pm0.02~M_\odot\,{\rm yr}^{-1}$. Given the very low $M_{\bigstar}$ of U13151 (see Section 3.1.2), these SFRs place the source well above the expected star-forming main sequence at these redshifts (see  \citealt{merida_probing_2025}), and firmly within the starburst regime \citep{caputi_star_2017, rinaldi_galaxy_2022, rinaldi_emergence_2024}, suggesting that it is undergoing a rapid phase of stellar-mass assembly with a stellar-mass doubling time of about $10.6\pm3.2\,\rm Myr$. Indeed, it is characterized by an extremely high ${\rm sSFR(H\alpha)}$ of $94\pm29\,{\rm Gyr}^{-1}$ placing it among the most intensely star-forming low-mass systems identified with JWST at $z\gtrsim10$,  potentially {\it the} most extreme known to date at $z\approx10$ (see Figure \ref{fig:ssfr}). We will discuss the incredible nature of such a starburst system later in this work and its possible implication for the Blue Monsters discovered by JWST (Section 4.1).

The extreme sSFR of U13151 may also have important implications for its dust content. Its ${\rm sSFR(H\alpha)}$ is $\approx4$ times higher than the critical value ${\rm sSFR}_{\bigstar}\approx25\,{\rm Gyr}^{-1}$ proposed by \citet{ferrara_super-early_2024} for the onset of a super-Eddington phase in compact early galaxies, and about $3$ times above the average expected at $z\approx9.8$ in their “fast-formation” model (${\rm sSFR}\propto(1+z)^{3/2}$). In their framework, radiation pressure from young stars can drive dust and gas to larger radii, reducing the effective dust optical depth. Thus, the dust-free condition inferred from H$\alpha$/H$\beta$ as well as from {\sc CIGALE} does not necessarily imply that dust has never formed in U13151. Indeed, \citet{langeroodi_rapid_2024} find that dust attenuation can emerge within $\lesssim30$ Myr of the onset of star formation, favoring rapid dust production by core-collapse supernovae (e.g., \citealt{gall_production_2011}). This means that U13151 could have had enought time before the observed epoch for earlier dust production. Its present extreme sSFR (see Figure~2 of \citealt{ferrara_super-early_2024}) could therefore prevent dust from accumulating, or redistribute it away from the compact UV-emitting regions, potentially aided by the shallow gravitational potential of this ultra-low-mass system. However, the current data cannot distinguish this scenario, and we do not attempt to discuss this further. Alternatively, following Occam's razor, U13151 may simply be undergoing its first major episode of star formation, with insufficient time for dust to build up efficiently within the system, as discussed by \citet{langeroodi_rapid_2024}.

The ratio between ${\rm SFR}_{\rm H\alpha}$ and ${\rm SFR}_{\rm UV}$ is commonly used to trace the strength of the most recent star-formation activity relative to that averaged over longer timescales; it is often referred to as the {\it burstiness} parameter ($\mathcal{B} = \rm SFR_{H\alpha}/SFR_{UV}$; e.g., \citealt{sparre_starbursts_2017,atek_star_2022}).  We find $\mathcal{B}=1.58\pm0.27$, implying that the SFR traced by H$\alpha$ is higher than that inferred from the UV continuum. Since H$\alpha$ is sensitive to the most recent star formation (within 10 Myr), while the UV continuum averages star formation over longer timescales (within 100 Myr), a ratio above unity suggests that U13151 has experienced an increase in its star-formation activity over the last few Myr \citep[e.g.,][]{atek_star_2022, faisst_recent_2019, iani_midis_2024-1, navarro-carrera_burstiness_2026}. Indeed, H$\alpha$ responds rapidly to the formation of short-lived massive stars, whereas the UV continuum evolves more slowly and can retain the imprint of earlier star formation. The measured ratio indicates that U13151 may be caught during a burst of star formation. 

\begin{figure}
    \centering
    \includegraphics[width=1.\linewidth]{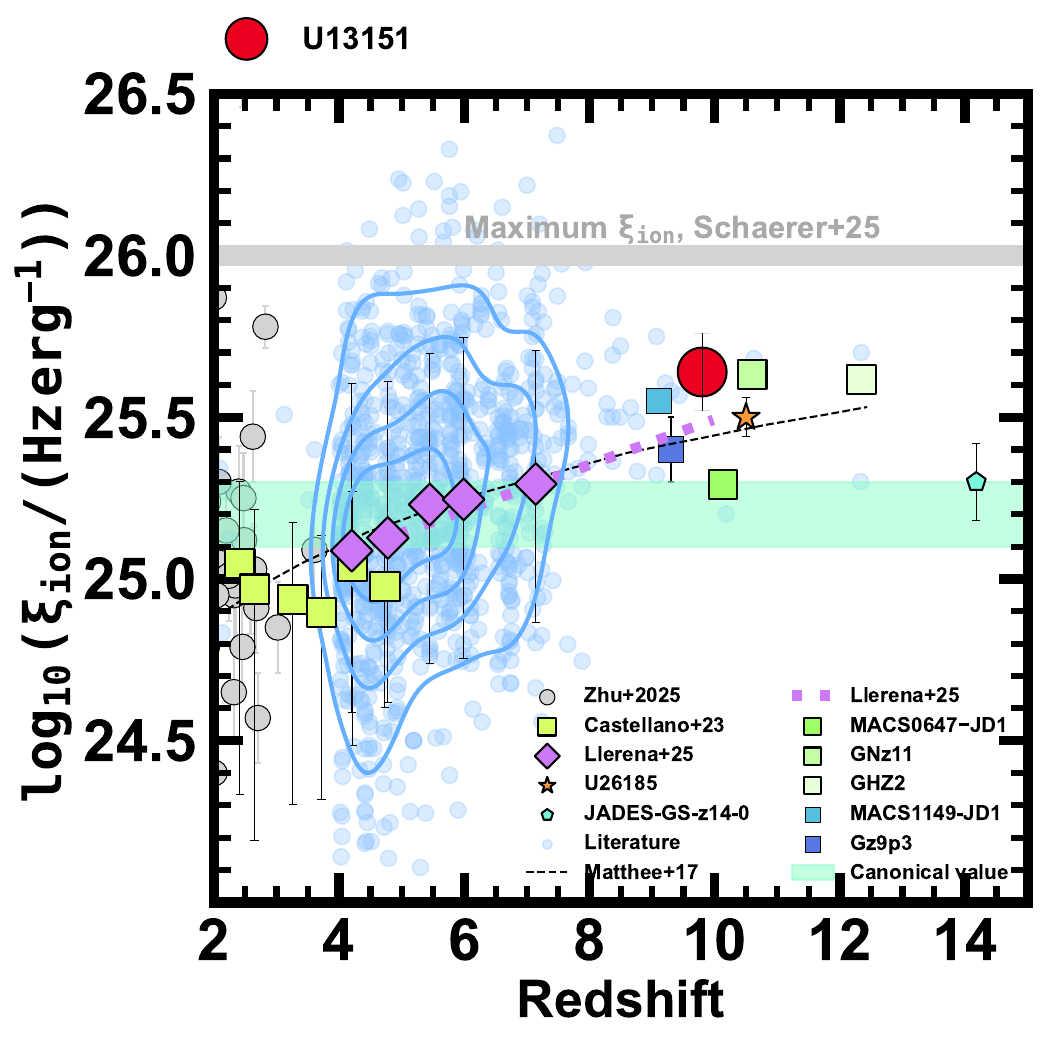}
    \caption{$\xi_{\rm ion}$ as a function of cosmic time. U13151 is shown as a red filled circle. Azure circles indicate measurements from the literature (both pre- and post-JWST; \citealt{stark_spectroscopic_2015, bouwens_lyman-continuum_2016, nakajima_hard_2016, marmol-queralto_evolution_2016, matthee_production_2017, stark_ly_2017, harikane_silverrush_2018, shivaei_mosdef_2018, faisst_recent_2019, lam_mean_2019, de_barros_greats_2019, emami_ionizing_2020, nanayakkara_reconstructing_2020, castellano_ionizing_2023, marques-chaves_extreme_2022, stefanon_high_2022, prieto-lyon_production_2023, bunker_jades_2023, alvarez-marquez_spatially_2024, rinaldi_midis_2024, roberts-borsani_between_2024, mascia_new_2024, lin_empirical_2024, zavala_luminous_2025}) as well as from surveys like JADES (\citealt{deugenio_jades_2024, curtis-lake_jades_2025, scholtz_jades_2025}), CEERS (\citealt{finkelstein_cosmic_2025}), and GLASS (\citealt{mascia_glass-jwst_2024}). For all these measurements, we also display their contours. Purple points show the median values from \citet{llerena_ionizing_2025}, together with their inferred evolution at $z\approx4\text{--}10$. At lower redshift, we show the evolution inferred by \citet{castellano_ionizing_2023}. We also include U26185, another galaxy observed with MIRI/LRS data, located at slightly higher redshift than U13151 (\citealt{alvarez-marquez_prisms_2026}). For comparison, we show other $z>10$ objects with direct $\xi_{\rm ion}$ measurements: GNz11 (\citealt{alvarez-marquez_insight_2025}), MACS0647-JD1 (\citealt{hsiao_jwst_2024}), MACS1149-JD1 (\citealt{alvarez-marquez_spatially_2024}), Gz9p3 (\citealt{bik_spatially_2026}), GHZ2 (\citealt{calabro_evidence_2024, zavala_luminous_2025}), and JADES-GS-z14-0 \citep{helton_ionizing_2026}. Both MACS0647-JD1 and JADES-GS-z14-0 lie much closer to the commonly assumed canonical value, $\log_{10}(\xi_{\rm ion}/{\rm Hz,erg^{-1}})=25.2\pm0.1$ \citep{robertson_new_2013}. Finally, we show, as a horizontal light gray line, the maximum $\xi_{ion}$ value inferred by stellar population synthesis models (\citealt{schaerer_observable_2025}).}
    \label{fig:xion_vs_z}
\end{figure}

Finally, since we have direct access to H$\alpha$, we estimate the ionizing photon production efficiency ($\xi_{\rm ion}$), which is a key quantity for quantifying the production of ionizing photons and therefore for understanding Cosmic Reionization. Following the method outlined in \citet{rinaldi_midis_2024}, we first derived the Lyman continuum escape fraction ($f_{\text{esc, LyC}}$) from the observed UV-$\beta$ slope using the prescription of \citet{chisholm_far-ultraviolet_2022}. We then computed $\xi_{\text{ion}}$ based on its dependence on $(1 - f_{\text{esc, LyC}})^{-1}$ (Equation 2 in \citealt{rinaldi_midis_2024}). We find $f_{\text{esc, LyC}} = 0.19 \pm 0.17$ and $\log_{10}(\xi_{\rm ion}/\text{Hz erg}^{-1}) = 25.64 \pm 0.12$\footnote{Even assuming $f_{\mathrm{esc,LyC}}\approx0\%$, the inferred $\xi_{\rm ion}$ would remain similar and consistently above the canonical value.}. The latter value lies well above the canonical estimate ($25.2\pm0.1$; \citealt{robertson_new_2013}) and is also slightly higher than the expected trend at such redshifts recently inferred by \citet{llerena_ionizing_2025}. Nonetheless, such elevated values are not unusual. 
In Figure~\ref{fig:xion_vs_z}, we show U13151 in the context of recent literature measurements up to $z\approx14$. 

The $\xi_{\rm ion}$ value we infer for U13151 is not unexpected; indeed, very high $\xi_{\rm ion}$ values are commonly found in strong H$\alpha$  emitters, i.e. systems with large rest-frame H$\alpha$ equivalent widths (EW$_0$; see Figure~5 of \citealt{rinaldi_midis_2024}). Indeed, the H$\alpha$ rest-frame equivalent width from the MIRI/LRS spectrum is constrained only as a lower limit, EW$_0({\rm H}\alpha)>938$~\AA\footnote{Following \citet{boyett_extreme_2024}, such a large EW would place the source among the extremely strong line emitters, which represent only a small fraction of the line-emitting population in the early Universe.}. We note that such high $\xi_{\rm ion}$ and EW(H$\alpha$) would be consistent with a very young stellar population ($\approx5$ Myr; see Figure 6 in \citealt{prieto-jimenez_spatially_2025}), as indeed we infer from CIGALE. Similarly, U13151 exhibits an exceptionally large EW$_{0}$ for [O\,{\sc iii}]$\lambda\lambda4959,5007$ ($> 1045$ \AA). Such large [O\,{\sc iii}] equivalent widths, as for H$\alpha$, are typically associated with high $\xi_{ion}$ (see Figure 10 from \citealt{llerena_ionizing_2025}). As such, U13151 and similar systems may contribute significantly to cosmic reionization. We will discuss the potential role of U13151-like galaxies in the early Universe later in this work (Section 4.3).

\subsubsection{Metallicity and abundance ratios}

\begin{figure}
    \centering
    \includegraphics[width=1.\linewidth]{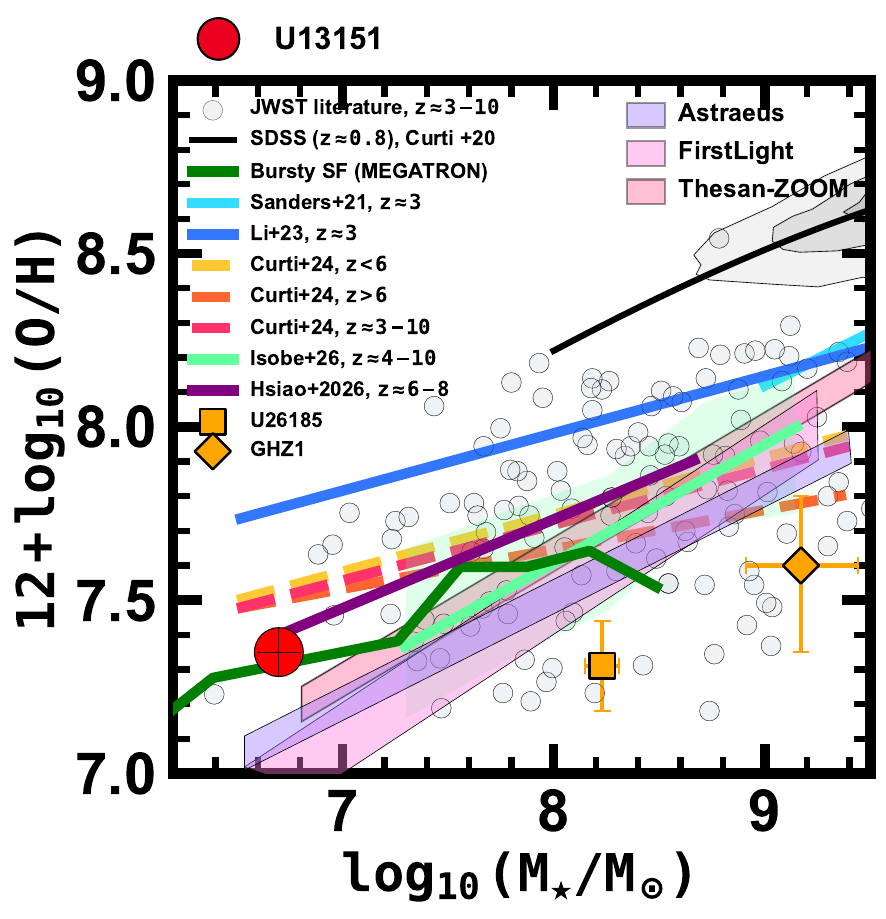}
    \caption{The stellar mass--gas-phase metallicity relation. U13151 is shown as a red filled circle. Gray contours denote recent JWST measurements at $z \approx 3$--10 from JADES \citep{curti_jades_2024}, CEERS \citep{nakajima_jwst_2023}, and SMACS J0723.3-7327 \citep{curti_chemical_2023}. Lower-redshift comparisons include the SDSS relation from \citet{curti_mass-metallicity_2020}, MOSDEF measurements at $z \approx 2$--3 from \citet{sanders_mosdef_2021}, and the low-mass $z \approx 3$ JWST/NIRISS relation from \citet{li_mass-metallicity_2023}. Dashed curves show the JADES MZR fits from \citet{curti_jades_2024}, split at $z=6$, together with the updated calibration of \citet{isobe_jades_2026}. We also show the recent results from the GLIMPSE survey (\citealt{hsiao_glimpse_2026}) at $z\approx6\text{--}8$. High-redshift model predictions from {\sc Astraeus} \citep{ucci_astraeus_2023}, {\sc FirstLight} \citep{langan_weak_2020}, and {\sc Thesan-zoom} \citep{mcclymont_thesan-zoom_2026} are also shown. Finally, we also show predictions from the {\sc MEGATRON} simulation in its bursty star-formation configuration (\citealt{katz_megatron_2025, choustikov_megatron_2026}). We also include U26185 and GHZ1 from the PRISMS survey (\citealt{alvarez-marquez_prisms_2026, perez-diaz_prisms_2026}).
}
    \label{fig:gas_phase_metallicity}
\end{figure}

Next we infer the gas-phase metallicity of U13151. To do so, we make use of different methodologies. First, we use the recent calibration at $z\approx1\text{--}10$ from \citealt{isobe_jades_2026} based on the recent JADES DR4 (\citealt{curtis-lake_jades_2025, scholtz_jades_2025}), which now extends this relation down to $\rm 12 +log(O/H)=7.0$. We use both the direct electron-temperature method and strong-line diagnostics. We note that the auroral [O\,{\sc iii}]$\lambda4363$ line is detected only at modest significance ($\rm SNR\approx2$).

We treat the direct-$T_{\rm e}$ metallicity as informative but uncertain. Assuming $n_{\rm e}=300~{\rm cm^{-3}}$ and no dust attenuation\footnote{We follow the fiducial $n_{e}$ value adopted in \citet{isobe_jades_2026}. The derived $T_{e}$-based abundance is only weakly sensitive to this assumption in the low-density regime. In particular, recent observations have shown that $n_e$ at $z\gtrsim10$ can vary substantially, spanning values from $\approx100$ to $\approx1000~{\rm cm}^{-3}$ (see, e.g., \citealt{abdurrouf_jwst_2024, larson_deep_2026}).}, motivated by the observed H$\alpha$/H$\beta$ ratio being consistent with the intrinsic Case-B value, we infer $T_{\rm e}({\rm O}^{2+})=18983_{-5160}^{+4063}$ K and $T_{\rm e}({\rm O}^{+})=16288_{-3612}^{+2844}$ K. This yields $12+\log({\rm O/H})_{\rm T_e}=7.44_{-0.17}^{+0.33}$, corresponding to $5.6^{+6.4}_{-1.8}\%\,Z_{\odot}$, adopting $12+\log({\rm O/H})_{\odot}=8.69$ (\citealt{asplund_chemical_2021}).

We also estimate the gas-phase metallicity using strong-line calibrations based on the available oxygen and neon indices. The combination based on $\log R_3$, $\tilde{R}$, $\log ({\rm O}_{32})$, and $\log({\rm Ne3O2})$ gives $12+\log({\rm O/H})_{\rm SL}=7.29^{+0.05}_{-0.05}$, or equivalently $Z_{\rm gas}=4.0^{+0.5}_{-0.4}\%\,Z_{\odot}$. We report these values in Table \ref{tab:spec}. 

Alternatively, we use the {\tt genesis-metallicity} code (\citealt{langeroodi_genesis-metallicity_2024}), which is based on a non-parametric combination of emission-line ratios. We infer $12+\log({\rm O/H})=7.27\pm0.04$ from the strong-line method, corresponding to $3.8\pm0.4\%\,Z_{\odot}$, and $12+\log({\rm O/H})=7.11\pm0.15$ from the direct $T_{\rm e}$ method, corresponding to $2.6^{+1.1}_{-0.8}\%\,Z_{\odot}$. The inferred electron temperatures are $T_{\rm e}({\rm O}^{2+})=23100\pm5800$ K and $T_{\rm e}({\rm O}^{+})=21300\pm600$ K.

Overall, these different methodologies consistently indicate that U13151 is a low-metallicity system, with a gas-phase metallicity spanning $12+\log({\rm O/H})\approx7.11$--$7.44$, corresponding to $\approx2.6$--$5.6\%\,Z_{\odot}$. Following the definition adopted by \citet{isobe_jades_2026}, U13151 falls within, or very close to, the regime of the extremely metal-poor galaxies (EMPGs) recently studied at high redshift with JWST (e.g., \citealt{isobe_jades_2026, trussler_out_2026}). Interestingly, U13151 falls on the expected trend derived by \citet{korber_glimpse-d_2026} between $12+\log({\rm O/H})$ and $M_{\rm UV}$, suggesting that fainter galaxies show, on average, lower metallicities. We remind the reader that also {\sc CIGALE} finds consistent results.

However, it is important to point out a caveat in our measurement based on the recent results of \citet{hsiao_optical_2026}, who showed that very high electron densities can substantially affect metallicity estimates in extremely metal-poor systems. At the NIRSpec/PRISM resolution, O\,{\sc iii}]$\lambda\lambda1660,1666$ cannot be resolved from He\,{\sc ii}, preventing an independent constraint on $n_{e}$ and motivating ultra-deep, higher-resolution spectroscopy.

In Figure \ref{fig:gas_phase_metallicity}, we place U13151 on the Mass--Metallicity Relation (MZR; \citealt{lequeux_chemical_1979, tremonti_origin_2004}), extending for the first time observational constraints on the MZR into the regime of extremely low-mass galaxies ($\log_{10}(M_\bigstar/M_\odot)<7$) at $z\approx10$. We caution that current observational constraints are not directly calibrated in this regime (i.e., ultra-low mass and $z\approx10$) and therefore require extrapolation. Within this caveat, U13151 appears slightly more metal enriched than expected for its stellar mass relative to recent JWST measurements at high redshift (\citealt{isobe_jades_2026,umeda_gas-phase_2026}) and to predictions from {\sc FirstLight}, {\sc Astraeus}, and {\sc Thesan-Zoom} (\citealt{langan_weak_2020,ucci_astraeus_2021,mcclymont_thesan-zoom_2026}), while remaining below the MZR inferred by \citet{curti_jades_2024}. Interestingly, its position is broadly consistent with the low-mass, bursty star-formation configuration of {\sc MEGATRON}, although these predictions are evaluated at $z\approx6$--8 \citep{katz_megatron_2025,choustikov_megatron_2026,hsiao_glimpse_2026}. In this scenario, stronger feedback drives enriched outflows and lowers the normalization of the MZR, while dense, turbulent star-forming gas can produce a relatively flat low-mass relation. U13151 might provide a valuable probe of the competition between rapid early enrichment and feedback-driven metal loss in the smallest galaxies $\approx500$ Myr after the Big Bang.

\begin{figure*}
    \centering
    \includegraphics[width=\textwidth]{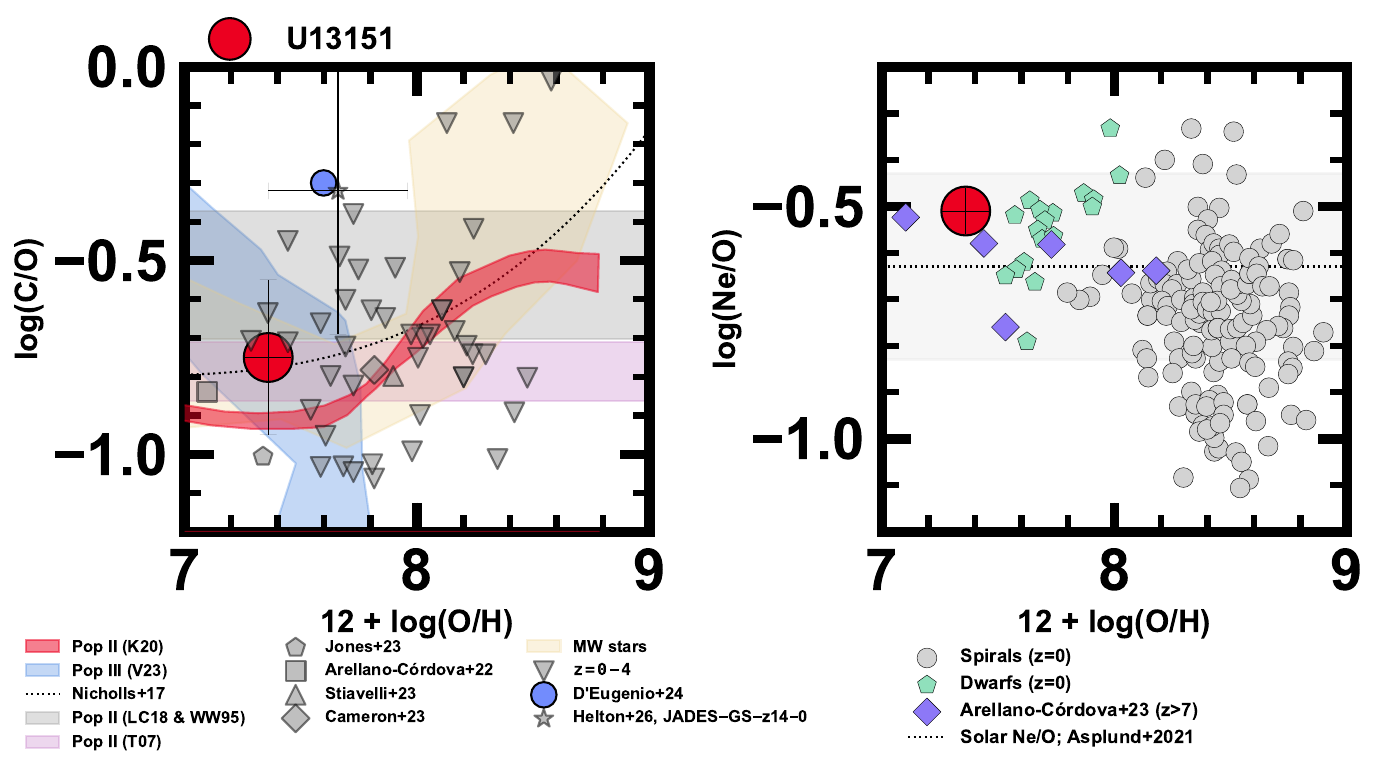}
    \caption{{\bf Left Panel:} C/O vs. O/H. We compare the observed abundance ratios with theoretical nucleosynthetic yields. The observational sample includes local stars (\citealt{gustafsson_origin_1999, akerman_evolution_2004, fabbian_co_2009, nissen_carbon_2014}) shown as a shaded yellow region to indicate the average parameter space they occupy, as well as low- and intermediate-z galaxies ($z\approx0\text{--}4$; \citealt{erb_physical_2010, christensen_gravitationally_2012, bayliss_physical_2014, james_testing_2014, stark_ultraviolet_2014, berg_carbon_2016, pena-guerrero_carbon_2017, senchyna_ultraviolet_2017, berg_window_2018, berg_chemical_2019, mainali_relics_2020, matthee_x-shooter_2021, iani_scrutiny_2023, citro_comprehensive_2024}) and high-z systems (\citealt{arellano-cordova_first_2022, cameron_nitrogen_2023, jones_early_2023, stiavelli_puzzling_2023}). Among the high-z objects, we show GS-z12 \citep{deugenio_jades_2024}, a $z=12.5$ object with an anomalously high C/O abundance (marked with a different color) as well as JADES-GS-z14-0 (though, in this case, error bars are quite large). For reference, we include the enrichment track from \citet{nicholls_abundance_2017}, Pop\,II supernova yields from \citet{tominaga_supernova_2007}, the Pop\,II abundance pattern predicted by \citet{kobayashi_origin_2020}, and the expected yields from metal-poor Type\,II supernovae based on \citet{woosley_evolution_1995} and \citet{limongi_presupernova_2018}. The Pop\,III yield region is taken from \citet{heger_nucleosynthesis_2010}, as adapted by \citet{vanni_characterizing_2023}.  {\bf Right Panel:} Ne/O as a function of O/H. We show $z\approx0$ low-metallicity dwarf galaxies and H\,\textsc{ii} regions in spiral galaxies from \citet{berg_chemical_2019} and \citet{berg_chaos_2020}, together with JWST measurements for $z>7$ galaxies from \citet{arellano-cordova_first_2022}. The black dashed line marks the solar abundance ratio, $\log_{10}({\rm Ne/O})=-0.63$ \citep{asplund_chemical_2021}, while the light gray shaded area represents the observed average scatter ($\approx0.2$~dex) in high-$z$ objects seen by \citet{arellano-cordova_first_2022}. U13151 is shown as a red filled circle in both panels.}
    \label{fig:abundances}
\end{figure*}

From the detected emission lines, we infer C/O and Ne/O following \citet{arellano-cordova_first_2022}. We show our measurements in Figure \ref{fig:abundances} along with recent literature measurements. For C/O, we find $\log_{10}({\rm C/O})=-0.84\pm0.18$. However, given the high ionization parameter of U13151 (see next section), we apply the photoionization-model ionization correction factor (ICF) from \citet{berg_chemical_2019}\footnote{We note that U13151 is even more metal poor than the grids adopted in their work.}, obtaining $\log_{10}({\rm C/O})=-0.75\pm0.19$; this is the value reported in Table \ref{tab:spec} and Figure \ref{fig:abundances}. A caveat in this measurement is that, at the spectral resolution of the PRISM, O\,{\sc iii}]$\lambda\lambda1660,1666$ cannot be robustly deblended from He\,{\sc ii}$\lambda1640$. A contribution from He\,{\sc ii} would lead to an overestimate of the O\,{\sc iii}] flux and, consequently, to an underestimate of C/O.

Both estimates indicate a low C/O ratio, corresponding to only $\approx25$--$35\%$ of the solar C/O ratio \citep{asplund_chemical_2021}. Together with the very low gas-phase metallicity, this suggests that U13151 is at an early stage of chemical enrichment. In this picture, oxygen is rapidly produced by short-lived massive stars and released predominantly through core-collapse supernovae (CCSNe), while the delayed carbon contribution from longer-lived stellar populations has not yet fully accumulated (\citealt{kobayashi_origin_2020}). Our measurements agree (within the uncertainties) with the expected evolution from \citet{nicholls_abundance_2017}. Such a enrichment dominated by the prompt massive-star channel is also consistent with the non detection  of N\,{\sc iv}]$\lambda1486$ and N\,{\sc iii}]$\lambda1750$ (upper limits provided in Table \ref{tab:spec}; in our case $\rm log_{10}(N/O)<-0.76$), in stark contrast with, e.g., GHZ2, a N-enhanced galaxy (\citealt{castellano_jwst_2024}). Nonetheless, it is worth noting that U13151 is consistent with both the expected Pop III track (\citealt{vanni_characterizing_2023}) and the Pop II track (\citealt{tominaga_supernova_2007}), highlighting that deeper observations will be required to better constrain the nature of its stellar population.

In contrast, we find $\log_{10}({\rm Ne/O})=-0.52\pm0.07$. Applying the ICF corrections from \citet{amayo_ionization_2021} gives $\log_{10}({\rm Ne/O})=-0.51\pm0.07$, which is the value reported in Table \ref{tab:spec} and Figure \ref{fig:abundances}. Although above the solar value ($\log_{10}({\rm Ne/O})_\odot=-0.63$; \citealt{asplund_chemical_2021}), our ICF-corrected value differs by only $\approx0.1$ dex and remains within the $\approx0.2$ dex range observed in high-redshift Ne/O measurements (see discussion in \citealt{arellano-cordova_first_2022}). Such behavior is expected in the presence of massive stars that heavily release elements via CCSNe.

The combination of an approximately solar Ne/O ratio and a strongly sub-solar C/O ratio suggests chemical enrichment dominated by CCSNe, with limited evidence for a significant delayed carbon contribution from older stellar populations. This picture is consistent with the bursty nature of U13151; the galaxy may be observed shortly after the onset of one of its first major episodes of star formation and, thus, chemical enrichment. 

\subsubsection{The extreme ISM condition of U13151}

The NIRSpec/PRISM spectrum of U13151 covers several key rest-frame optical emission lines, including [O\,{\sc ii}]$\lambda\lambda$3727,3729, [Ne\,{\sc iii}]$\lambda$3869, H$\delta$, H$\beta$, and [O\,{\sc iii}]$\lambda\lambda$4959,5007. Although some of these emission lines are detected at relatively modest significance, their measurements still allow us to investigate the physical conditions of the interstellar medium (ISM) through a set of complementary emission-line diagnostics, namely O32, R23, Ne3O2, and Ne3O2Hd (see Table \ref{tab:emission_ratios}). To interpret these diagnostics, we compare the measured ratios with {\sc CLOUDY} photoionization models computed following the prescriptions adopted by \citet{calabro_near-infrared_2023} and \citet{calabro_evidence_2024}\footnote{The latest version of these models (for both AGNs and SFGs) is available at \url{https://github.com/Anthony96}.} (see also \citealt{rinaldi_deciphering_2025}). We refer the reader to these works for a full description of the model assumptions and explored parameter space.

\begin{figure*}
    \centering
    \includegraphics[width=\textwidth]{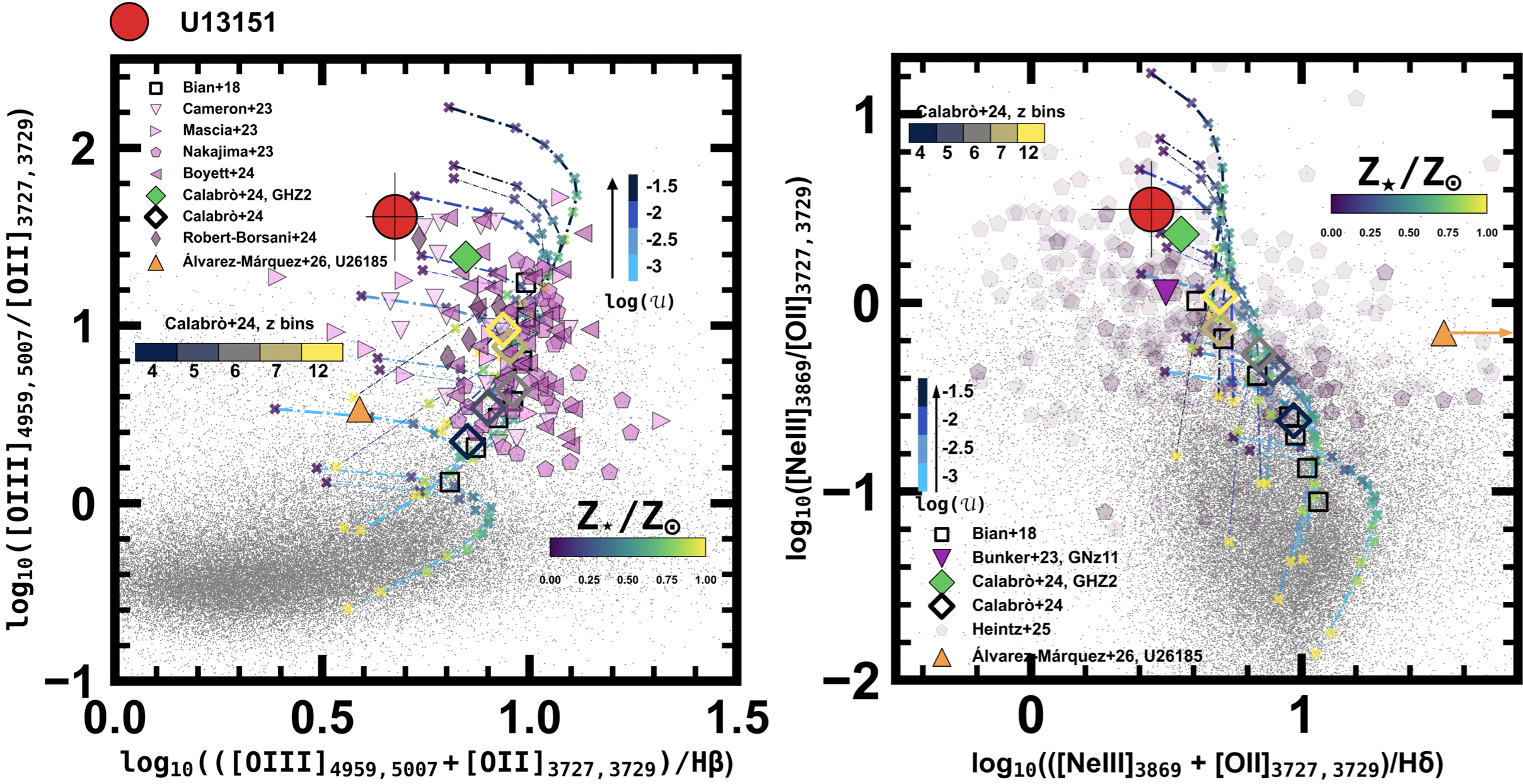}
    \caption{{\bf Left Panel:} O32--R23. {\bf Right Panel:} Ne3O2Hd--Ne3O2. U13151 is marked with a red filled circle. In both diagrams, we compare our measurements with {\sc CLOUDY} photoionization models spanning electron densities of $n_{\rm e}=10^{2}$, $10^{3}$, and $10^{4}$ cm$^{-3}$, shown with progressively thicker symbols/curves; ionization parameters of $\log_{10}(\mathcal{U})=-3$, $-2.5$, $-2$, and $-1.5$, indicated by different blue shades; and gas-phase metallicities from $0.05,Z_{\odot}$ to $Z_{\odot}$, color-coded with a viridis scale and marked by ``x'' symbols. Further details on the model grid are provided in \citet{rinaldi_deciphering_2025}, which is based on \citet{calabro_evidence_2024}. As reference samples, we include low-redshift star-forming galaxies from SDSS \citep{york_sloan_2000, kauffmann_stellar_2003} and nearby analogues of high-redshift galaxies from \citet{bian_direct_2018}. We also show literature measurements for galaxies at $z\approx4$--$12$ from \citet{bunker_jades_2023, cameron_jades_2023, mascia_closing_2023, nakajima_jwst_2023, boyett_extreme_2024, calabro_evidence_2024, calabro_evolution_2024, roberts-borsani_between_2024, heintz_jwst-primal_2025, alvarez-marquez_prisms_2026}. 
}
    \label{fig:ism}
\end{figure*}

The R23--O32 plane provides a useful view of the evolving ionization state of galaxies across cosmic time. As shown by \citet{calabro_evidence_2024}, galaxies move toward increasingly high-ionization conditions with increasing redshift, from the locus occupied by local SDSS galaxies to the region populated by sources at $z\approx4$--12. In Figure~\ref{fig:ism} (left panel), we compare U13151 with recent high-redshift NIRSpec measurements from \citet{cameron_jades_2023, mascia_closing_2023, nakajima_jwst_2023, boyett_extreme_2024, calabro_evidence_2024, calabro_evolution_2024, roberts-borsani_between_2024}, together with local SDSS galaxies \citep{york_sloan_2000, kauffmann_stellar_2003}. We also include the low-redshift metal-poor, high-ionization sample of \citet{bian_direct_2018}, commonly used as a local analog population for high-redshift galaxies.

U13151 occupies the upper part of the R23--O32 diagram, placing it among the most highly ionized systems currently identified at these redshifts. Its location is even more extreme than that of GHZ2 (see \citealt{calabro_evidence_2024}). More generally, U13151 follows the locus traced by other EoR galaxies, supporting the idea that its nebular gas is characterized by very intense ionizing radiation fields. When compared with the star-forming-galaxy photoionization grids, U13151 overlaps with a relatively narrow region of parameter space. The remaining degeneracy is mainly associated with the assumed electron density, with viable models spanning $n_{\rm e}=10^{2}$--$10^{4}\,\mathrm{cm}^{-3}$. However, across this range, the models consistently favor moderately high ionization parameters ($\log_{10}(\mathcal{U})\approx -2$) and low $Z_{\rm gas}$ ($\lesssim10\%\,Z_{\odot}$). 

We have a direct estimate of its $Z_{\rm gas}$ (previous section) and can characterize the ionization state of U13151. For the latter, we can do so from the O32 ratio by using the calibration of \citet{papovich_clear_2022} for CLEAR galaxies at $1.1<z<2.3$ \citep{simons_clear_2023}. We obtain $\log_{10}(\mathcal{U})=-2.28\pm0.24$. Independent rest-frame UV diagnostics from \citet{mingozzi_classy_2022}, based on C\,\textsc{iii}], O\,\textsc{iii}], and C\,\textsc{iv}, give an average value of $\log_{10}(\mathcal{U})=-1.88\pm0.15$. The $\mathrm{Ne3O2}$--$\log_{10}(\mathcal{U})$ relation from \citet{witstok_assessing_2021} gives a slightly higher value, $\log_{10}(\mathcal{U})=-1.63\pm0.23$. Although the individual diagnostics span a broad range, from ionization parameters typical of star-forming galaxies ($\log_{10}(\mathcal{U})\approx-2.3$) to substantially more extreme values reaching $\log_{10}(\mathcal{U})\approx-1.6$, they suggest that the ISM may host regions characterized by very high ionization. We therefore adopt the median of the three estimates, with a conservative uncertainty that accounts for both measurement scatter and calibration systematics, obtaining $\log_{10}(\mathcal{U})=-1.90\pm0.35$. Overall, the position of U13151 in the R23--O32 diagram indicates very extreme ISM conditions, in line with what has been found for other galaxies at such high redshift, including GHZ2 \citep{calabro_evidence_2024}.

We further examine the ISM properties of U13151 using the Ne3O2Hd--Ne3O2 diagram, shown in the right panel of Figure~\ref{fig:ism}. As in the R23--O32 comparison, we include local SDSS galaxies and the metal-poor, high-ionization analogs from \citet{bian_direct_2018}. U13151 again falls in the region occupied by high-redshift galaxies, including recent JWST measurements such as those reported by \citet{heintz_jwst-primal_2025}, and follows the redshift evolution identified by \citet{calabro_evolution_2024} over $z\approx4$--12. Compared to the R23--O32 plane, this diagnostic shows reduced degeneracy with respect to the photoionization models, providing a more direct indication that U13151 hosts a highly ionized, very metal-poor ISM.

Altogether, the extreme position of U13151 in both diagnostic diagrams indicates that its nebular gas is exposed to an unusually intense ionizing radiation field, as also derived independently from {\sc CIGALE}. Recently, high values of O32 and Ne3O2 have been associated to high ionization parameters, elevated electron densities, and intense star formation concentrated within compact regions (\citealt{reddy_jwstnirspec_2023, topping_metal-poor_2024, cleri_rubies_2026}). Its location at the extreme end of these diagrams may be the proxy of a very young and intense burst of stellar assembly (as also independently derived from e.g. SED fitting and burstiness), similar to the increasingly extreme ISM conditions observed in galaxies at $z\gtrsim9$ (\citealt{tang_jwst_2026}).

\subsubsection{Does U13151 host an AGN?}

Given the extreme ISM conditions of U13151, we investigate whether it shows evidence for AGN activity. This is particularly challenging at high redshift, where classical optical diagnostics become ambiguous. Baldwin, Phillips \& Terlevich (BPT; \citealt{baldwin_classification_1981}) diagrams were calibrated locally and are strongly metallicity dependent. In the early Universe, young, metal-poor, and highly ionized star-forming galaxies can shift toward the AGN locus, while AGN narrow-line regions may overlap with the SFG sequence (e.g., \citealt{nakajima_diagnostics_2022}). An important example is GS\_3073 at $z\approx5.55$, which shows broad H$\alpha$ and H$\beta$ emission, but it lies in the SFG region of the BPT diagram \citep{ubler_ga-nifs_2023}. These caveats become relevant for U13151, whose very low metallicity, high ionization parameter, and extreme ISM conditions place it close to the EMPG regime.

For this reason, we focus on the [O\,\textsc{iii}]$\lambda4363$-based diagnostic proposed by \citet{mazzolari_new_2024}. This has been proposed as a better suited diagram to search for AGN signatures in the recent literature, and has been shown to be effective in the faint AGN regime now being uncovered by JWST, including the newly population of Little Red Dots (LRDs; e.g., \citealt{
matthee_little_2024}) and Little Blue Dots (LBDs; e.g., \citealt{ brazzini_little_2026}).

\begin{figure*}
    \centering
    \includegraphics[width=1\linewidth]{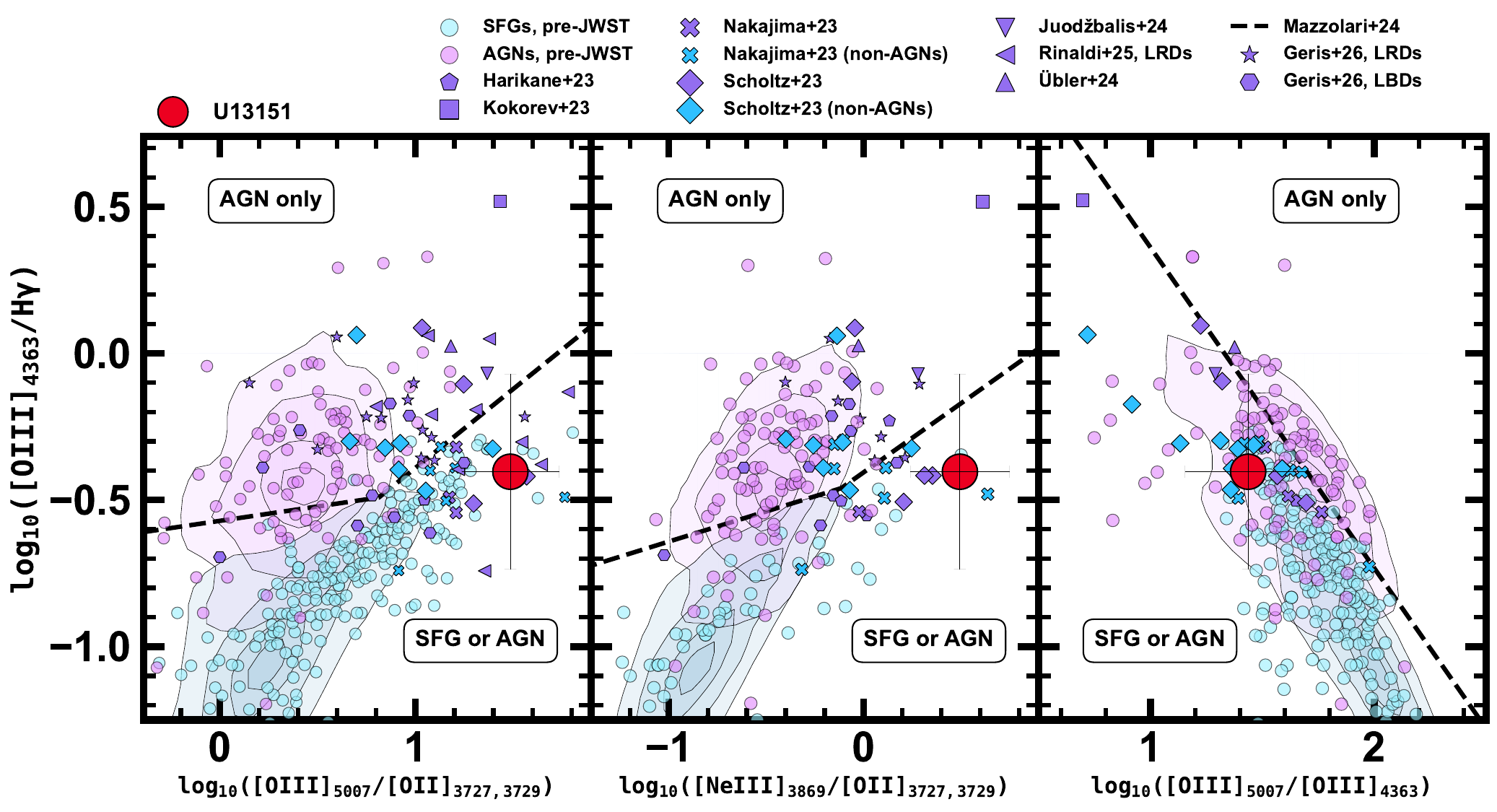}
    \caption{{\bf Left Panel:} O32--O3Hg. {\bf Middle Panel:} Ne3O2--O3Hg. {\bf Right Panel:} O33--O3Hg. U13151 is marked with a red filled circle. In each panel, the black dashed lines show the empirical boundaries introduced by \citet{mazzolari_new_2024} to separate AGN-dominated systems from star-forming galaxies. The shaded contours show the distribution of local galaxies from SDSS \citep{abazajian_seventh_2009}, used as a low-redshift reference sample. Filled symbols indicate pre-JWST measurements of both SFGs and AGNs, mostly drawn from low-redshift studies \citep{seyfert_nuclear_1943, izotov_chemical_2006, berg_direct_2012, amorin_extreme_2015, perna_x-raysdss_2017, yang_blueberry_2017, yang_ly_2017, izotov_j08114730_2018, dors_chemical_2020, armah_chemical_2021, pustilnik_xmp_2021}. We also include recent JWST measurements from \citet{harikane_jwstnirspec_2023, nakajima_jwst_2023, scholtz_jades_2023, ubler_ga-nifs_2024, juodzbalis_dormant_2024}, which include both star-forming and AGN candidates. Moreover, we show the sample of LRDs and LBDs from \citet{rinaldi_not_2025} and \citet{geris_little_2026}.
}
    \label{fig:u13151_line_ratios_oiii_hg}
\end{figure*}

In Figure~\ref{fig:u13151_line_ratios_oiii_hg}, we compare U13151 with local AGNs and SFGs, as well as recent JWST samples of high-redshift AGNs and  SFGs (including also LRDs and LBDs). Across all three diagnostic diagrams, U13151 lies in the extreme region of the parameter space, close to where several high-redshift AGNs are found, including both type-I and type-II AGNs, as well as LRDs and LBDs. However, the relatively large uncertainties on the U13151 line ratios prevent a firm classification.  Though, one can still think about two scenarios. On the one hand, given the extreme conditions of U13151, even these diagnostics become hard to use to classify such objects, similarly to what has been observed in BPT-like diagrams \citep{ubler_ga-nifs_2023}. On the other hand, U13151 may host a composite system, with both star formation and a (perhaps weak) AGN contributing to its ionizing radiation field. It is also important to note that different regions within this system may exhibit distinct line ratios, reflecting spatial variations in their physical conditions and metallicity, as observed in other similarly high-redshift systems (e.g., Gz9p3; \citealt{bik_spatially_2026}).

\begin{figure}
    \centering
    \includegraphics[width=1.\linewidth]{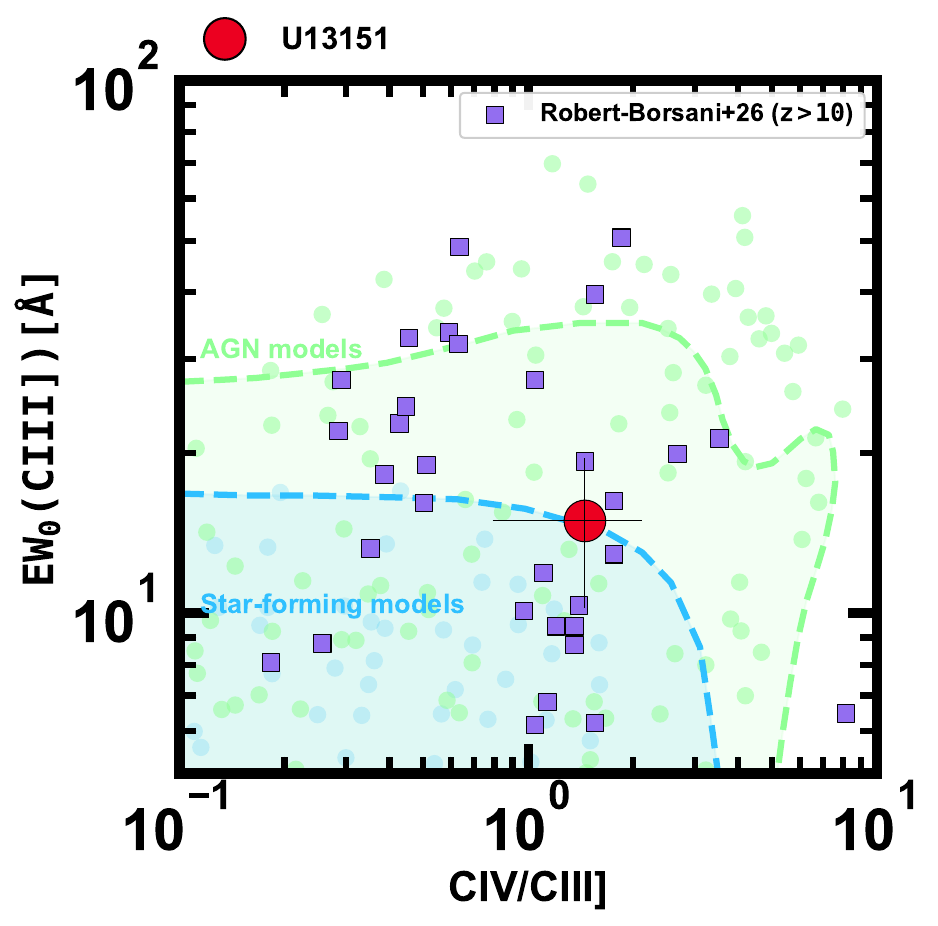}
    \caption{C\,{\sc iv}/C\,{\sc iii} vs. $\rm EW_{0}$(C\,{\sc iii}]) diagnostic. Star-forming and AGN models from \citet{nakajima_diagnostics_2022} are shown (as circles along with their contours enclosing $>95\%$ of the model data points). U13151 is shown as a red filled circle, which is close to the boundary between star-forming galaxies and AGNs though with high error bars. This might suggest that, while an intense star formation episode remains a viable explanation as a source of ionization, a weak AGN contribution cannot be excluded, and deeper high resolution data are required to shed light on it.}
    \label{fig:ciii_civ_diagnostic}
\end{figure}

As a final diagnostic, we examine the C\,{\sc iii}] equivalent width and the C\,{\sc iv}/C\,{\sc iii}] ratio. We measure $\rm EW_{CIII]} = 14.9\pm4.7\,\AA$ and C\,{\sc iv}/C\,{\sc iii}] $= 1.45\pm0.63$. We stress that these measurements remain explorative, and deeper data are required for a more robust interpretation. Nevertheless, they provide an independent route to investigate the source of ionization in U13151. In Figure \ref{fig:ciii_civ_diagnostic}, we place U13151 in the context of the recent measurements presented by \citet{roberts-borsani_jwst_2026} at $z\gtrsim10$, together with the photoionization models from \citet{nakajima_diagnostics_2022}. In this case, U13151 lies close to the boundary between the regions occupied by star-forming and AGN models, suggesting that both scenarios remain viable. While the current data do not allow us to draw a firm conclusion, this diagnostic again leaves open the possibility that U13151 might be a composite system with, perhaps, a weak AGN contribution. 

Overall, deeper, higher-resolution UV spectroscopy will be crucial to finally identify the dominant source of ionization in this peculiar system and to assess whether U13151 could host an accreting intermediate-mass black hole (IMBH).

\section{Discussion}\label{sec4}

\subsection{Rapid stellar-mass assembly through compact clumps in the early Universe}

U13151 is an intrinsically ultra-faint, ultra-low-mass galaxy caught during an exceptionally intense burst of star formation, {\it likely} its first major episode of star formation. Its lensing-corrected stellar mass is only $\log_{10}(M_\bigstar/M_\odot)=6.68\pm0.11$, while its stellar age of $5.5\pm2.8$ Myr indicates that its emission is dominated by a very young stellar population\footnote{Even accounting for outshining by a young stellar population, the inferred $M_\bigstar$ would increase by only a factor of approximately two on average and would remain exceptionally low for such a high redshift (e.g., \citealt{hamed_there_2026}).}. From the MIRI/LRS detection of H$\alpha$, we infer ${\rm SFR}_{\rm H\alpha}=0.45\pm0.08\,M_\odot\,{\rm yr}^{-1}$, corresponding to stellar-mass doubling time of $M_\bigstar/{\rm SFR}_{\rm H\alpha}\approx10.6\pm3.2$ Myr. Although this estimate neglects stellar mass loss and should not be interpreted as evidence that the current, likely transient burst will persist (\citealt{sparre_starbursts_2017,madau_cosmic_2014}), it highlights the extreme intensity of the ongoing episode, perhaps {\it the most} intense starburst system observed so far at $z\gtrsim10$. This interpretation is further supported by ${\rm SFR}_{\rm H\alpha}/{\rm SFR}_{\rm UV}=1.58\pm0.27$ and by the very high equivalent widths of [O\,{\sc iii}]$\lambda\lambda4959,5007$ and H$\alpha$, both indicative of nebular emission dominated by a very young population of massive stars. Interestingly, U13151 shares several properties with the recently identified population of N-enhanced galaxies, including 1) its extreme compactness, 2) intense star formation episode, and 3) very young stellar population (e.g., \citealt{topping_metal-poor_2024}). Such conditions have been proposed to facilitate rapid, localized enrichment by CNO-processed material from massive and Wolf--Rayet stars (e.g., \citealt{rusakov_diverse_2026}). However, U13151 shows no evidence for N-enhanced, and with its abundance patterns (i.e., relatively low C/O and approximately normal Ne/O) being consistent with prompt enrichment dominated by massive stars through CCSNe. Such a stark contrast may suggest that similar boundary conditions observed in N-enhanced systems might not map strictly one-to-one onto N enhancement, and that other mechanisms could be at play.

Remarkably, the rest-frame UV morphology offers a direct indication of where this rapid growth is occurring. Our {\sc lenstronomy} decomposition of the F150W image identifies five compact clumps contributing approximately $18\%$, $9\%$, $50\%$, $13\%$, and $10\%$ of the modeled UV flux, with the brightest component (3) alone accounting for about half of the total UV emission (Figure \ref{fig:f150w_decomposition}). We measure an observed effective radius of $R_{\rm e}=122.4_{-7.2}^{+7.9}$ pc after correcting for lensing. This is consistent with the compact sizes measured for other galaxies at comparable redshifts (\citealt{ono_morphological_2025}). Using this $R_e$, we infer $\Sigma_{\rm SFR(UV)}\approx3\,M_\odot\,{\rm yr}^{-1}\,{\rm kpc}^{-2}$ and $\Sigma_{\rm SFR(H\alpha)}\approx4.8\,M_\odot\,{\rm yr}^{-1}\,{\rm kpc}^{-2}$. Taken at face value, these values are not exceptional for galaxies at $z\approx10$ and are broadly consistent with theoretical expectations and recent observational trends (e.g., \citealt{yung_semi-analytic_2022,calabro_evolution_2024}). However, essentially all of the detected UV luminosity is associated with the five compact components, and no dominant smooth component is observed. If we distribute the total UV SFR according to the F150W flux fractions and adopt the lensing-corrected F150W PSF radius ($R_{\rm e}\approx30$ pc) as an upper limit on individual clump sizes, we get conservative lower limits of $\Sigma_{\rm SFR,UV,clump}\gtrsim4.6$--$25.6\,M_\odot\,{\rm yr}^{-1}\,{\rm kpc}^{-2}$. These may substantially underestimate the instantaneous local star-formation intensity, since 1) the UV averages over longer timescales than H$\alpha$, 2) U13151 shows ${\rm SFR}({\rm H\alpha})/{\rm SFR}({\rm UV})$ ratio above unity, and 3) the true clump sizes may be considerably smaller than our PSF-based upper limit (see e.g. \citealt{adamo_bound_2024} where they infer $R_{e}\approx 1\,\rm pc$ for a single clump). If the unresolved H$\alpha$ emission is concentrated in only a subset of the UV-bright clumps, the instantaneous $\Sigma_{\rm SFR}$ in those regions would be higher still. Taken together with the very low inferred clump masses, down to a few $10^{5}\,M_\odot$ (assuming that the UV flux is a good tracer of $M_{\bigstar}$), this suggests that U13151 is a multi-clump, bursty system in which a substantial fraction of its $M_{\bigstar}$ is being assembled rapidly within compact star-forming regions.

U13151 is not unique, though. Other examples exist in the recent literature. For instance, the Cosmic Gems Arc at $z=9.625$ is resolved into five massive stellar clusters with characteristic masses of $\approx10^{6}\,M_\odot$, effective radii close to $1$ pc, and $\Sigma_\star\approx10^{5}$--$10^{6}\,M_\odot\,{\rm pc}^{-2}$ (\citealt{adamo_bound_2024,messa_jwst_2026}). These clusters contribute a substantial fraction of the host UV light and stellar mass and appear to have formed during a short burst, making them plausible proto-globular clusters. Similarly, the Firefly Sparkle at $z=8.304$ contains ten compact components with masses of $\approx10^{5.3}$--$10^{5.8}\,M_\odot$ and radii of only a few parsecs, collectively accounting for approximately $49$--$57\%$ of the galaxy's stellar mass (\citealt{mowla_formation_2024}). Such systems may represent an early phase of galaxy assembly in which dense stellar structures form in gas-rich, turbulent environments and subsequently contribute to the emerging stellar body through migration, stripping, expansion, or disruption (see discussion in \citealt{mowla_formation_2024}). Similar systems are now being identified at even higher redshift. BulletArc-z11 ($z=11.10$) contains three compact components with sizes of $\lesssim10$--$25$ pc that together contribute approximately half of the inferred stellar mass, with stellar surface densities reaching $\gtrsim10^{5}\,M_\odot\,{\rm pc}^{-2}$ (\citealt{bradac_star_2025}). Even more remarkable is the Misty Moons system at $z\approx11$--$12$, which contains compact components with radii of $\approx10$--$70$ pc and masses of order $10^{7}\,M_\odot$ that collectively account for $\gtrsim80\%$ of the inferred stellar mass (\citealt{nakane_venus_2025}).   Although these systems span a wide range in clump size, mass, and compactness, they point toward a common picture in which, at least for some low-mass ultra-faint systems during the first few hundred Myr of cosmic time, the preferred mode of stellar mass assembly is through bursty, multi-clump star formation.

Assuming $M_{\bigstar}$ follows the UV light, the clumps have $M_{\bigstar}\approx10^{5.7}-10^{6.4}\,M_\odot$. Given $R_{\rm e}<30$ pc, this implies $\Sigma_{\bigstar}\gtrsim10^{2.0}-10^{2.5}\,M_\odot\,{\rm pc}^{-2}$. These values approach those observed in compact globular clusters and galactic nuclei in the low-redshift Universe (e.g., \citealt{mccrady_kinematic_2003, mccrady_super_2007, leroy_forming_2018, ostlin_dynamical_2007}), suggesting a possible connection between the dense stellar structures forming at very high redshift and compact stellar systems observed in the local Universe. If the clumps subsequently evolve and merge on timescales set by the internal dynamics of the system, this assembly could proceed rapidly (see e.g. \citealt{dekel_formation_2009}). Indeed, recent simulations on clumpy high-redshift galaxies show that clumps might migrate inward and coalesce in a few tens of Myr (\citealt{nakazato_merger-driven_2024}). Perhaps, U13151 is being observed during a phase in which $M_{\bigstar}$ is being assembled simultaneously within several dense clumps that could later merge into a more centrally concentrated system.

\begin{figure}
    \centering
    \includegraphics[width=1.\linewidth]{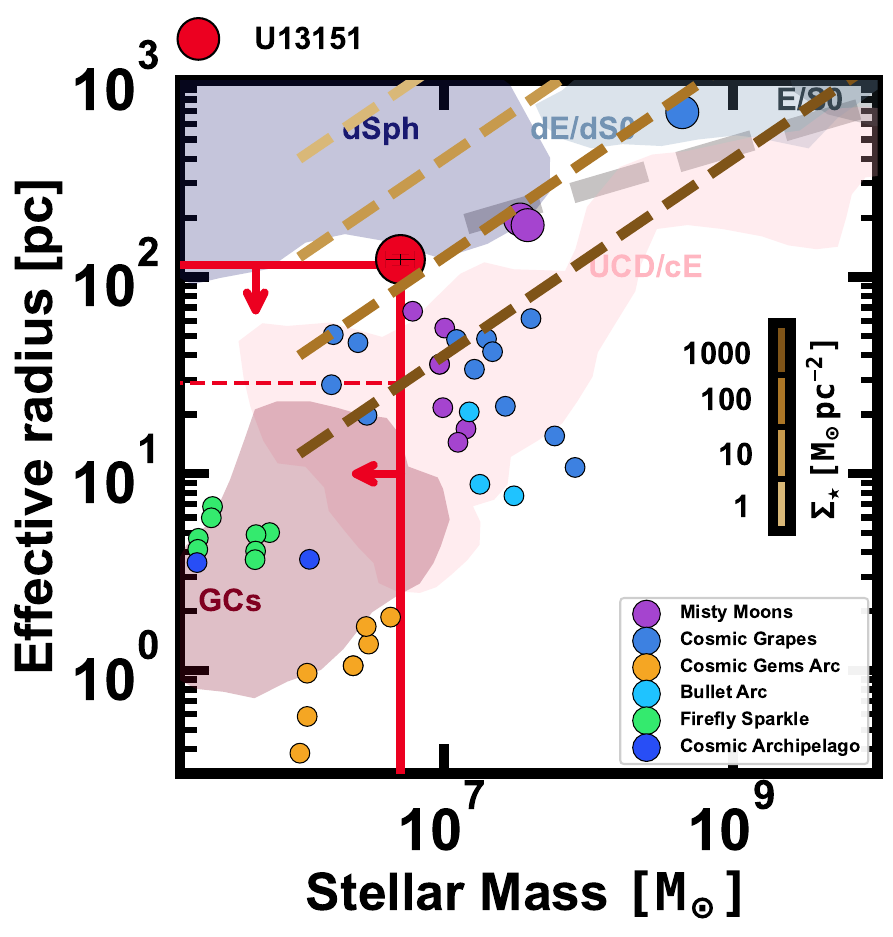}
    \caption{$M_{\bigstar}$ vs. effective-radius. We show U13151 along with other compact systems in the early Universe: the Misty Moons at $z\approx11$--12 (\citealt{nakane_venus_2025}), Cosmic Grapes at $z\approx6.1$ (\citealt{fujimoto_primordial_2025}), Cosmic Gems Arc at $z\approx9.625$ (\citealt{adamo_bound_2024, bradley_unveiling_2025, messa_jwst_2026}), BulletArc-$z11$ at $z\approx11.1$ (\citealt{bradac_star_2025}), Firefly Sparkle at $z\approx8.325$ (\citealt{mowla_formation_2024}), and Cosmic Archipelago at $z\approx6.145$ (\citealt{messa_anatomy_2025}). For the Misty Moons and Cosmic Grapes systems, larger symbols mark their integrated quantities. Shaded regions show local systems from dwarf spheroidals to globular clusters (\citealt{norris_aimss_2014}), while the gray relation shows the $z\approx4$--10 size--mass relation from \citet{langeroodi_ultraviolet_2023}; brown lines indicate constant surface densities of 1, 10, 100, and 1000 $\rm M_\odot\,pc^{-2}$. U13151 lies at lower $M_{\bigstar}$ than most comparable high-$z$ systems, even allowing for  outshining effects (\citealt{hamed_there_2026}). Red solid lines indicate the approximate location of its individual clumps, assuming UV-light fractions trace (1 to 1) stellar-mass fractions, corresponding to masses of a few $10^{5}$--$10^{6},M_\odot$. The dashed red line marks the clump-size upper limit from the F150W HWHM ($\approx0.025$\arcsec), placing them in the regime of compact star-forming superclusters and massive stellar clusters. 
}
    \label{fig:size_mass}
\end{figure}

In Figure \ref{fig:size_mass}, we show U13151 along the systems described above, noting that its integrated $M_{\bigstar}$ inferred from SED fitting is lower than that of most comparable systems studied in the recent literature. Even if our $M_{\bigstar}$ were underestimated by a factor of two because of outshining by the dominant young stellar population (e.g., \citealt{hamed_there_2026}), the total stellar mass of U13151 would still remain below those inferred for similar systems such as Misty Moons, BulletArc-z11, the Cosmic Grapes, and the Cosmic Gems Arc. It is also important to highlight another important property of U13151.

When placed in context, U13151 may represent a particularly {\it early} manifestation of clump-driven stellar-mass assembly. Its $M_{\bigstar}$ is comparable to, or even lower than, the combined mass contained in the compact structures of some of the systems discussed above, while its stellar age is comparable to the duration of a single intense star-forming episode. Indeed, at its present SFR, U13151 could reproduce its existing $M_{\bigstar}$ within only a few Myr. In more evolved galaxies, star-forming clumps generally contribute to the growth of an already substantial underlying stellar population (e.g., \citealt{guo_clumpy_2015}), as is also the case for young massive clusters in nearby dwarf galaxies (e.g., \citealt{weisz_acs_2011}). In U13151, by contrast, the distinction between the clumps and the galaxy itself may be less clear. If most of its young stellar population is concentrated in these components, the clumps may constitute the primary sites where the main body of the galaxy is being assembled, similar to the Firefly Sparkle system (\citealt{mowla_formation_2024}); see also discussion in \cite{nakane_venus_2025}.

In this regime, clump formation and galaxy growth may occur nearly simultaneously, with the galaxy initially building up through the formation and subsequent merging of dense stellar structures (\citealt{mayer_situ_2025}). Such a scenario is consistent with the high gas fractions, pressures, densities, and $\Sigma_{\rm SFR}$ expected in early galaxies, which should favour star formation in gravitationally bound structures (see \citealt{adamo_star_2020} and references therein).

Several mechanisms could generate such a clumpy configuration. For instance, in a gas-rich system, gravitational instability may fragment a nascent disk into dense star-forming structures, particularly when the gas fraction is high and the dynamical time is short (e.g., \citealt{dekel_formation_2009, ceverino_high-redshift_2010}). Alternatively, fragmentation may begin within cold accreting streams before a stable disk has formed, producing compact stellar systems distributed along the inflowing gas (e.g., \citealt{dekel_formation_2009, mandelker_cold_2018}). Turbulence-driven density fluctuations may also promote locally high star-formation efficiencies, while interactions or mergers between small progenitors could produce multiple spatially distinct components. In the FirstLight simulations, such clumpy morphologies are predominantly merger-driven, with young off-centre clumps forming from tidally perturbed gas and subsequently coalescing on timescales of a few tens of Myr (\citealt{nakazato_merger-driven_2024}); similar merger-driven scenarios have also been proposed by \citet{ginzburg_origin_2025}.
More recently, \citet{ferrara_clumpy_2026} suggested that the emergence of compact stellar clumps in super-early galaxies may itself be regulated by feedback. In their framework, dense clumps formed through gravitational instability can remain embedded in their natal gas until radiation pressure becomes strong enough to clear the surrounding material, allowing the young stellar component to emerge. For metal-poor systems, this clearing is expected to be particularly efficient through resonantly trapped Ly$\alpha$ photons, before dust-mediated infrared pressure becomes dominant at later stages of enrichment.

Regardless of their origin, compact clumps can strongly regulate the subsequent evolution of such a low-mass system as U13151. If the clumps reach high local star-formation efficiencies, some may remain gravitationally bound and survive as long-lived stellar clusters (e.g., \citealt{kruijssen_formation_2012,grudic_when_2018,polak_massive_2024}). Others may expand or dissolve after gas removal, depositing their stars into a more diffuse galactic component (e.g., \citealt{goodwin_gas_2006}). 

With the growing number of compact, clumpy systems identified at very high redshift, it is tempting to interpret objects such as U13151 and, for instance, the Cosmic Gems as being observed at different stages of a common, strongly time-variable evolutionary cycle. In very low-mass galaxies, the shallow gravitational potential may cause star formation to respond rapidly to the accumulation and removal of cold gas. Gas inflow or local collapse could trigger a short episode of highly efficient and spatially clustered star formation, followed by feedback-driven heating, dispersal, or expulsion of the remaining gas. The system would then enter a temporarily suppressed or post-burst phase until the gas cools or is replenished; this would induce highly stochastic SFHs (e.g., \citealt{stinson_breathing_2007, sparre_starbursts_2017, mcclymont_thesan-zoom_2025}). Within this tentative picture, the Cosmic Gems may represent a phase in which compact clusters remain visible while star formation in the surrounding region has already declined (\citealt{messa_jwst_2026}), while U13151 may be observed at a much earlier and more active stage, during an intense burst before feedback has substantially reduced its nebular emission.

Finally, in the hierarchical picture of galaxy formation, the first galaxies are expected to assemble within relatively low-mass dark matter halos, progressively growing through gas accretion, star formation, and mergers into more massive systems. This naturally raises the possibility that the exceptionally UV-bright Blue Monsters discovered by JWST at $z\gtrsim10$ may have originated from much smaller and intrinsically fainter building blocks at even earlier cosmic times. In the framework proposed by \citet{ferrara_eventful_2024}, sufficiently compact galaxies can enter a super-Eddington regime when their sSFR exceeds a metallicity-dependent critical value, potentially driving radiation-driven dusty outflows. For the very low metallicity inferred for U13151 ($\approx3$--$6\%\,Z_\odot$), this corresponds to ${\rm sSFR}_\bigstar\approx23$--$47\,\rm Gyr^{-1}$, well below its observed ${\rm sSFR(H\alpha)}\approx94\,{\rm Gyr}^{-1}$. Remarkably, the observed physical conditions of U13151 are already consistent with the strongly bursty, super-Eddington regime proposed in these models, although at a much lower $M_{\bigstar}$ than typically associated with Blue Monsters; U13151 itself might represent a {\it late} pre-Blue Monster.

Interestingly, \citet{kokorev_glimpse_2025} showed that intrinsically faint ($M_{\rm UV}\approx-17$) galaxy candidates at $z\approx16$ could evolve into the exceptionally UV-bright galaxies observed at $z\approx10\text{--}14$, provided sufficiently efficient star formation (see their Figure 7). Under the simple assumption that U13151 is representative of similarly faint galaxies at much higher redshifts, it is tempting to speculate that such systems could follow similar evolutionary pathways. In this picture, the extreme physical conditions observed in U13151 may provide a glimpse into the earliest stages of rapid stellar-mass assembly, potentially preceding the emergence of the Blue Monster population. Although highly speculative, this possibility further emphasizes the importance of exploring the faintest galaxies at the highest redshifts. In this regard, future ultra-deep MIRI/LRS observations in strong-lensing fields will be crucial for probing the ultra-faint galaxy population at $z>10$ and tracing Blue Monsters back to their possible earlier evolutionary stages.

\subsection{U13151 as a progenitor of a Milky Way-like galaxy seen $\approx500$ Myr after the Big Bang?}
In the previous section, we showed how U13151 is rapidly assembling possibly most, if not all, of its $M_{\bigstar}$ predominantly within compact clumps. We also argued that U13151-like systems could represent the early progenitors of Blue Monsters if observed at much higher redshifts. In this section, instead, we explore an alternative evolutionary scenario in which U13151-like systems could represent the early building blocks of much more massive galaxies observed in the present-day Universe (e.g., our Milky Way or M31). It is important to stress, however, that tracing the evolution of such systems from these early cosmic epochs to the present day is highly uncertain and degenerate. U13151 could remain a low-mass satellite or eventually merge into a much larger system, losing its identity as an individual galaxy. Nonetheless, it is tempting to investigate how its observed properties compare with predictions from models calibrated to reproduce the formation and evolution of present-day MW/M31-like galaxies.

\begin{figure}
    \centering
    \includegraphics[width=1.\linewidth]{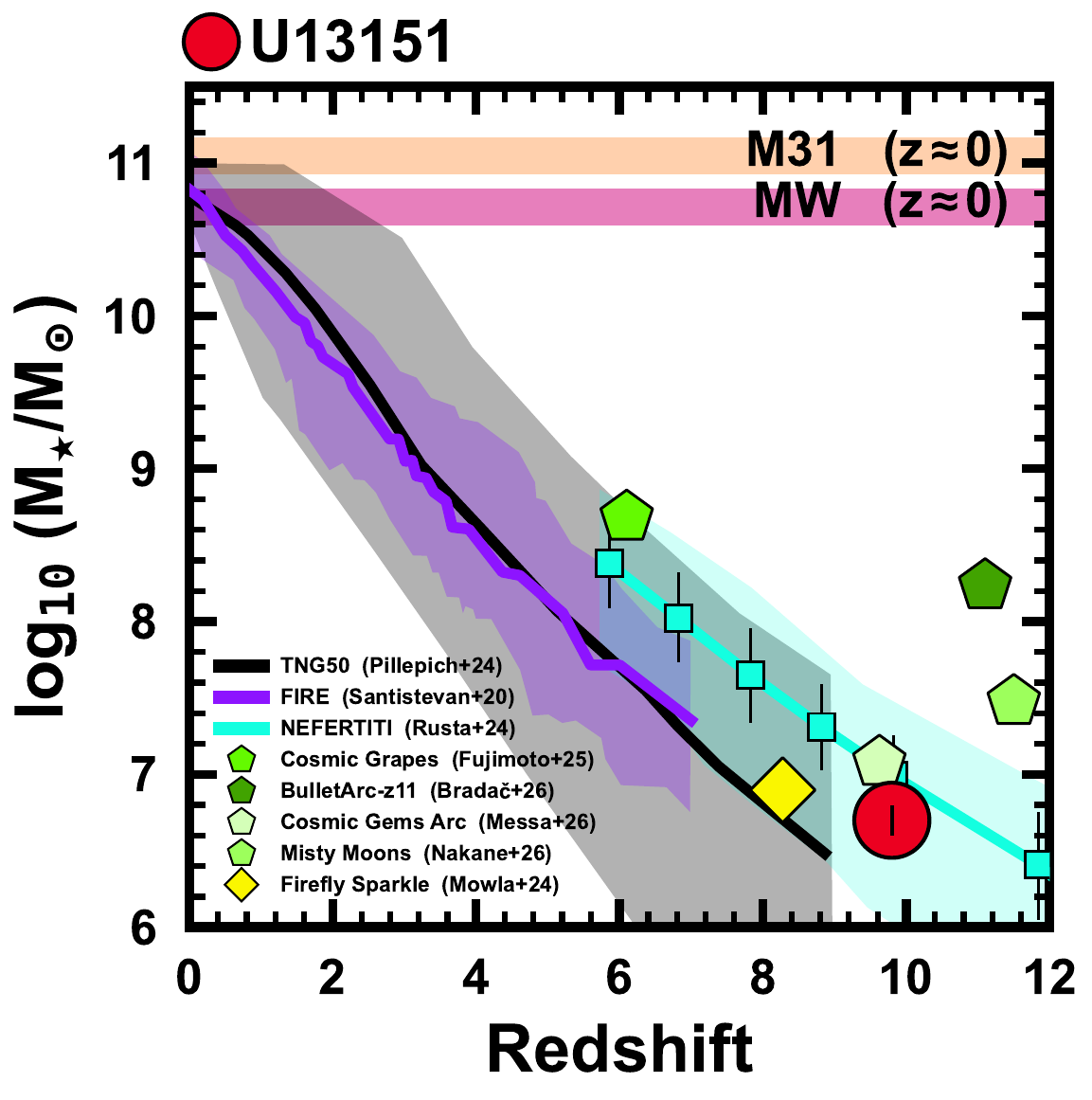}
    \caption{We compare U13151 with the progenitors of Milky Way analogues identified in the {\sc TNG50} (\citealt{pillepich_milky_2024}), {\sc FIRE} (\citealt{santistevan_formation_2020}), and {\sc NEFERTITI} (\citealt{rusta_linking_2024}) simulations. U13151 lies at a substantially higher redshift than the range directly probed by the {\sc TNG50} and {\sc FIRE} predictions, which are calibrated up to $z\approx7\text{--}8$, but remains within approximately $1\sigma$ of their extrapolated evolutionary tracks. By contrast, U13151 lies near the lower envelope of the {\sc NEFERTITI} predictions, closely matching the lowest-$M_{\bigstar}$ and lowest-metallicity Milky Way progenitor identified by \citet{rusta_linking_2024}. We also show other systems broadly similar to U13151 and the Firefly Sparkle: the Misty Moons at $z\approx11\text{--}12$ (\citealt{nakane_venus_2025}), the Cosmic Grapes at $z\approx6.1$ (\citealt{fujimoto_primordial_2025}), the Cosmic Gems Arc at $z\approx9.625$ (\citealt{adamo_bound_2024, bradley_unveiling_2025, messa_jwst_2026}), and BulletArc-$z11$ at $z\approx11.1$ (\citealt{bradac_star_2025}). 
}
    \label{fig:progenitor}
\end{figure}

To discuss this possibility, we follow the recent analyses of the Firefly Sparkle at $z=8.304$ presented by \citet{mowla_formation_2024} and \citet{rusta_linking_2024}, and place U13151 within possible assembly histories of MW/M31 analogs (see Figure \ref{fig:progenitor}). Most predictions for MW progenitors have so far been calibrated primarily up to $z\approx7$--$8$, including those based on the {\sc TNG50} (\citealt{pillepich_milky_2024}) and {\sc FIRE} simulations (\citealt{santistevan_formation_2020}). 

\citet{mowla_formation_2024} showed that the Firefly Sparkle lies within $\approx1\sigma$ of the expected $M_{\bigstar}$ of a Milky Way progenitor at $z\approx8.3$, which they interpret as an object caught in an exceptionally early phase of assembly where they both observe compact star clusters and a diffuse stellar arc. When extrapolating the MW progenitor assembly tracks from both {\sc TNG50} and {\sc FIRE} toward lower $M_{\bigstar}$, U13151 remains broadly consistent with these trends, potentially suggesting that it could trace an earlier stage of the sequence proposed for the Firefly Sparkle. Interestingly, \citet{rusta_linking_2024} extended this framework to earlier epochs, following the assembly of a Milky Way-like system back to $z\approx14\text{--}15$ and applying it to the Firefly Sparkle. In their merger trees, the main branch reaches $\log_{10}(M_{\bigstar}/M_\odot)\approx6.8$--$7.0$ by $z\approx10$ (though a substantial scatter is observed; see their Figure 5). Their models also predict a much broader population of lower-mass building blocks spanning wide ranges of stellar mass and metallicity, which are subsequently accreted onto or merged with the main branch.

Within their framework, we find that the inferred $M_\bigstar$ of U13151 is fully consistent with the population of lower-mass progenitor systems that are later accreted onto or merged with the main branch. In particular, U13151 closely resembles some of the lowest-mass Firefly analogs identified by \citet{rusta_linking_2024}, for which the model predicts $\log_{10}(M_{\bigstar}/M_\odot)=6.26\pm0.10$ to $6.64\pm0.09$ and $Z_{\rm gas}\approx4\%\,Z_\odot$ to $\approx10\%\,Z_\odot$ (see their Table 1). Thus, within their framework, it is reasonable that U13151 {\it could be} a low-mass progenitor of a Milky Way-like galaxy. However, U13151 has a substantially higher SFR than the simulated progenitors of \citet{rusta_linking_2024}, further highlighting the exceptionally intense and rapid stellar-mass assembly that U13151 is undergoing at the moment of observations.

\subsection{The role of U13151-like systems during Cosmic Reionization}

It is now well established that faint, low-mass galaxies undergo bursty star formation (\citealt{sparre_starbursts_2017}). For instance, \citet{perry_prevalence_2025} found that, at $z\approx1\text{--}7$, galaxies with $\log_{10}(M_\bigstar/M_\odot)\lesssim8.5$ are $30\pm1\%$ more likely than higher-mass systems to be observed while undergoing or fading out of a recent burst of star formation ($\lesssim10\text{--}50$ Myr; see Figure 7 in \citealt{navarro-carrera_burstiness_2026}), with little-to-no dependence on cosmic time (see Figure 5 from \citealt{perry_prevalence_2025}). More generally, \citet{perry_prevalence_2025} infer that, across all redshifts and $M_{\bigstar}$, $73\pm4\%$ of the galaxies in their sample have H$\alpha$-to-UV ratios inconsistent with a constant SFH (e.g., \citealt{simmonds_bursting_2025, navarro-carrera_burstiness_2026}). 

Motivated by this evidence, in this section we want to explore the role that low-$M_{\bigstar}$ galaxies undergoing a U13151-like phase (i.e. a short-lived phase of extreme star formation in an ultra-faint, ultra-low-mass galaxy) could play during Cosmic Reionization. We remind the reader that an anti-correlation between $\xi_{\rm ion}$ and $M_{\bigstar}$ has been reported in both observations (e.g., \citealt{rinaldi_midis_2024, llerena_ionizing_2025}) and theoretical models (e.g., \citealt{seeyave_first_2023}), whereby lower-$M_{\bigstar}$ galaxies exhibit higher $\xi_{\rm ion}$. 

To explore this possibility, we construct a simple toy model based on the characteristic burst timescale proposed by \citet{stiavelli_how_2026} for galaxies at $z>5$. We define low-$M_{\bigstar}$ systems as galaxies with $\log_{10}(M_{\bigstar}/M_{\odot})\lesssim8.5$, following the division adopted by \citet{perry_prevalence_2025}. They find that $53\pm5\%$ of galaxies with $\log_{10}(M_\bigstar/M_\odot)\lesssim8.5$ lie above the equilibrium H$\alpha$-to-UV ratio, corresponding to $\mathcal{B}>1$. This choice is also broadly consistent with the increasing prevalence and scatter of bursty star formation toward lower stellar masses reported by \citet{simmonds_bursting_2025} and \citet{navarro-carrera_burstiness_2026} at high redshifts. We caution, however, that most of the current studies are not complete at $\log_{10}(M_{\bigstar}/M_{\odot})\lesssim8\text{--}8.5$ at any given time, and the behavior of the burstiness distribution below these $M_{\bigstar}$ remains still poorly constrained.

Based on the framework of \citet{stiavelli_how_2026}, galaxies grow through discrete star-forming episodes separated by a characteristic interval of $\Delta t\approx100$ Myr (see also \citealt{ceverino_firstlight_2019}). Their preferred model assumes approximately linear stellar-mass growth, with successive bursts contributing comparable stellar masses. Because \citet{stiavelli_how_2026} do not report a formal uncertainty on $\Delta t$, we treat it as a nuisance parameter and adopt a sensitivity prior of $\Delta t=100\pm25$ Myr, truncated to $50<\Delta t/{\rm Myr}<150$\footnote{We use \texttt{emcee} to propagate the uncertainties.}.
For simplicity, we assume that the most extreme U13151-like phase lasts $t_{\rm extreme}=10\pm3$ Myr\footnote{Motivated by the young stellar age inferred from {\sc CIGALE} and the stellar-mass doubling time inferred from H$\alpha$.}. Under this toy-model assumption, the corresponding duty cycle is $f_{\rm U13151}=t_{\rm extreme}/\Delta t$, yielding $f_{\rm U13151}\approx10^{+4}_{-3}\%$. We regard this estimate as optimistic because it implicitly assumes that every low-mass galaxy undergoes one U13151-like phase during each characteristic burst cycle.


For simplicity, we assign galaxies outside the U13151-like state a reference ionizing photon production efficiency of $\log_{10}(\xi_{\rm ion}/{\rm Hz\,erg^{-1}})=25.2\pm0.1$ and an escape fraction of $f_{\rm esc}=10\%$ (\citealt{robertson_new_2013}). For galaxies observed in the U13151-like state, we adopt $\log_{10}(\xi_{\rm ion}/{\rm Hz\,erg^{-1}})=25.64\pm0.12$, as inferred for U13151. We stress that this two-state description is deliberately simplistic\footnote{Because the U13151-like and reference states are assumed to arise from the same underlying UV luminosity density ($\rho_{\rm UV}$), the common factor $\rho_{\rm UV}$ cancels when computing their relative contribution to the escaping ionizing emissivity. The resulting fraction therefore depends only on the assumed duty cycle and on the values of $\xi_{\rm ion}$ and $f_{\rm esc}$ adopted for the two states.}; observations at high redshifts show a range of possible $\xi_{\rm ion}$ values (see \citealt{llerena_ionizing_2025} for reference). We do not attempt to explore the continuum of intermediate ionizing efficiencies between the two adopted states. Since $f_{\rm esc}$ for U13151 is not well constrained ($f_{\rm esc}=19\pm17\%$ using the relation from \citealt{chisholm_far-ultraviolet_2022}), we repeat the calculation assuming $f_{\rm esc}=1\%$, $5\%$, $10\%$, $15\%$, and $20\%$ for U13151-like systems. For $f_{\rm esc}=1\%$, U13151-like phases contribute $3.0^{+2.1}_{-1.2}\%$ of the escaping ionizing emissivity produced by the selected low-mass population. For $f_{\rm esc}=5\%$, $10\%$, $15\%$, and $20\%$, the corresponding contributions are $13.4^{+7.9}_{-5.1}\%$, $23.6^{+11.5}_{-8.3}\%$, $31.7^{+13.1}_{-10.3}\%$, and $38.2^{+13.7}_{-11.6}\%$, respectively. Therefore, U13151-like phases could contribute substantially to the escaping ionizing photon budget of low-mass galaxies, potentially accounting for $\approx24\%$ even for an escape fraction of $f_{\rm esc}=10\%$.

We stress that these fractions do not represent the contribution of U13151-like galaxies to the total ionizing photon budget of the Universe, but their relative contribution to the escaping ionizing emissivity of the low-$M_{\bigstar}$ population. In any case, this very simple exercise suggests that U13151-like phases {\it could} contribute non-negligibly to the ionizing output of low-mass galaxies without being either long-lived or numerically dominant. We also note that the inferred contribution depends strongly on the $f_{\rm esc}$, and could have huge impacts on the total ionizing contribution (e.g., see the case of U37126 at $z\approx10.255$; \citealt{marques-chaves_prisms_2026}). In the latter case, such sources would be dominating the ionizing photon budget, yet constraining $f_{\rm esc}$ remains a difficult task. Finally, an additional source of uncertainty arises from the assumed duration of the bursty phase represented by U13151.

Recently, \citet{jecmen_glimpse_2026} found that galaxies with $-14<M_{\rm UV}<-18$ may contribute approximately $60\%$ of the total ionizing photon budget. If the mass--luminosity relation ($M_{\rm UV}$--$M_\bigstar$) inferred at $z>6$ in the recent literature (e.g., \citealt{bhatawdekar_evolution_2019, furtak_how_2021, atek_jwst_2023}) is adopted as a first-order indicator of $M_{\bigstar}$, this luminosity range is expected to be largely populated by galaxies with $\log_{10}(M_\bigstar/M_\odot)\lesssim8.5$. Although, U13151 lies toward the bright end of this luminosity interval and, given its exceptionally low $M_\bigstar$, represents an extreme object within this broader population, it illustrates the potentially important role of intrinsically faint galaxies during Cosmic Reionization.

Finally, U13151 raises the possibility that the ionizing efficiency is particularly enhanced during the first episodes of star formation within similar systems. If the anti-correlation between $\xi_{\rm ion}$ and $M_\bigstar$ holds (\citealt{seeyave_first_2023}), ultra-faint, ultra-low-$M_{\bigstar}$ systems undergoing their first major burst of star formation may briefly act as exceptionally efficient factories of ionizing photons. In this {\it very simplified} picture, U13151-like systems would undergo short “cosmic-firework” phases during which a substantial fraction of their $M_{\bigstar}$ and ionizing radiation is produced.   Therefore, a systematic census of such ultra-faint, ultra-low-mass systems at $z>6$ is crucial to assess accurately their contribution to EoR, with lensing fields playing an important role in this effort.

\section{Summary and Conclusions}\label{sec5}

In this work, we analyzed, for the first time, the rest-frame UV-to-optical emission of U13151, an ultra-faint, ultra-low-mass galaxy observed only $\approx500$~Myr after the Big Bang previously reported by \citet{roberts-borsani_nature_2023} and \citet{fujimoto_uncover_2024}. The combination of NIRCam, NIRSpec, and MIRI (imaging and LRS) observations, together with the magnification ($\mu=12.8\pm0.7$) provided by gravitational lensing, offers us a rare glimpse into what may be the {\it first breath} of a galaxy in the early Universe, before a fully assembled galaxy has emerged. We summarize our main findings below:
\begin{itemize}
 
    \item The UV emission of U13151 can be resolved into five compact clumps; all of the detected UV emission is associated with these clumps, with no prominent smooth underlying component identified. An underlying host {\it could be} still present, though current data cannot allow us to detect it. Overall, U13151 is extremely compact, in line with recent studies at high redshifts (\citealt{ono_morphological_2025}).

    \item The combination of NIRCam and MIRI photometry with NIRSpec/PRISM and MIRI/LRS spectroscopy enables us to tightly constrain the physical properties of U13151. After correcting for lensing magnification, we infer an intrinsically very low stellar mass of $\log_{10}(M_{\bigstar}/M_{\odot})=6.68\pm0.11$ for the young stellar component and $M_{\rm UV}=-17.44\pm0.07$ mag. Given its non-detection in MIRI/F1000W, we cannot constrain the presence of an underlying older stellar population, for which we derive a $3\sigma$ upper limit of $M_{\bigstar,\rm old}<1.64\times10^7\,M_\odot$. Both SED fitting and the H$\alpha$/H$\beta$ Balmer decrement indicate virtually no dust, consistent with its very blue UV continuum slope ($\beta=-2.64\pm0.15$). {\sc CIGALE} returns a remarkably young stellar age of $5.5\pm2.8$ Myr. Finally, {\sc CIGALE} yields a high ionization parameter ($\log_{10}(\mathcal{U})=-1.57\pm0.35$), consistent with estimates obtained from multiple emission-line calibrations (\citealt{papovich_clear_2022, mingozzi_classy_2022, witstok_assessing_2021}), suggesting extreme ISM conditions comparable to or more extreme than those of other high-redshift systems (e.g., GHZ2; \citealt{calabro_evidence_2024}).

    \item MIRI/LRS enables, for the first time, the direct detection of H$\alpha$ in an ultra-faint system at $z\approx10$. In particular, we infer ${\rm SFR}_{\rm H\alpha}=0.45\pm0.08\,M_\odot\,{\rm yr}^{-1}$, approximately twice the UV-based value of ${\rm SFR}_{\rm UV}=0.29\pm0.02\,M_\odot\,{\rm yr}^{-1}$. The resulting $\mathcal{B}(\equiv{\rm SFR}_{\rm H\alpha}/{\rm SFR}_{\rm UV})=1.58\pm0.27$ indicates a recent increase in star-formation activity. At its current SFR, U13151 could build its entire stellar mass in only $\approx10.6\pm3.2$ Myr, highlighting the very high intensity of its burst relative to predictions at similar redshifts and firmly placing it in the starburst regime (\citealt{caputi_star_2017, rinaldi_galaxy_2022, rinaldi_emergence_2024}). Interestingly, such a stellar mass doubling time agrees, within the error bars, with the stellar age derived from {\sc CIGALE}, which suggests that possibly most (if not all!) of its $M_{\bigstar}$ has been assembled during its current star-formation episode.

    \item U13151 shows extremely strong emission lines, with  EW$_0$([O\,{\sc iii}]$\lambda\lambda4959,5007)>1045$~\AA\, and EW$_0$(H$\alpha)>938$~\AA\;(from NIRSpec/PRISM and MIRI/LRS), comparable to similarly extreme line emitters observed during EoR (\citealt{boyett_extreme_2024,rinaldi_midis_2023}). From H$\alpha$, we infer $\log_{10}(\xi_{\rm ion}/{\rm Hz\,erg}^{-1})=25.64\pm0.12$, well above the canonical value (\citealt{robertson_new_2013}), indicating that U13151 is a highly efficient producer of ionizing photons.

    \item We infer a very low gas-phase metallicity of $12+\log_{10}({\rm O/H})\approx7.11\text{--}7.44$, corresponding to $\approx3\%\text{--}6\%\,Z_\odot$, slightly more metal enriched compared to expectations for galaxies of similarly low $M_{\bigstar}$ at high redshift (e.g., \citealt{curti_jades_2024, isobe_jades_2026, umeda_gas-phase_2026}) and in agreement with results from SED fitting. We also infer a low ${\rm C/O}$ ratio ($\log_{10}({\rm C/O})=-0.75\pm0.19$), together with an approximately solar ${\rm Ne/O}$ ratio ($\log_{10}({\rm Ne/O})=-0.51\pm0.07$). This is consistent with U13151 being observed during one of its first major episodes of chemical enrichment, dominated by the prompt contribution of massive stars, before delayed enrichment channels have had sufficient time to contribute significantly. Interestingly, U13151 shares several physical properties with much brighter systems (e.g., GHZ2). Though, it does not show clear evidence of N enhancement. This stark contrast suggests that the usual conditions associated with N-enhanced galaxies (e.g., compactness, low metallicity, and intense star formation) may not necessarily map one-to-one onto enhanced N abundances.

    \item In the [O\,{\sc iii}]$\lambda4363$ diagnostic diagrams of \citet{mazzolari_new_2024}, U13151 lies close to the demarcation line between AGNs and SFGs and overlaps the region occupied by some JWST-selected AGNs, including LRDs. Similarly, the C\,{\sc iv}/C\,{\sc iii}] versus EW$_0$(C\,{\sc iii}]) diagram places it close to the demarcation between AGNs and SFGs. However, the large uncertainties in both cases prevent a firm classification. If anything, U13151 appears to be dominated by star formation (given its extreme ISM conditions), with at most a weak AGN contribution; future ultra-deep, high-resolution spectroscopy will be crucial for determining whether U13151 hosts an AGN at such low $M_{\bigstar}$.

    \item U13151 may represent a particularly early stage of clump-driven galaxy assembly, in which essentially all of its $M_{\bigstar}$ is currently being built directly within compact clumps. Its similarity to other strongly lensed systems suggests that this spatially concentrated mode of growth may be common among low-mass galaxies in the early Universe. Given its extreme properties, including its very high sSFR, low metallicity, and young stellar age, with a very simple assumption we speculate that U13151-like systems at earlier epochs could represent progenitors of the Blue Monsters observed with JWST at $z>10$ (see also discussion in \citealt{kokorev_glimpse_2025}). This highlights the urgent need to push MIRI/LRS to observe ultra-faint objects at much earlier times to fully uncover how the first galaxies formed and evolved.

    \item U13151 closely resembles the Firefly Sparkle at $z\approx8.3$ (\citealt{mowla_formation_2024}) and lies along the expected stellar-mass growth of MW/M31 progenitors in {\sc TNG50} and {\sc FIRE} (\citealt{pillepich_milky_2024, santistevan_formation_2020}). Similarly, its $M_{\bigstar}$ and metallicity are also consistent with the low-mass building blocks in \citet{rusta_linking_2024}, again suggesting that U13151 could represent one of the early components later accreted into MW/M31-like systems.

    \item Using a simple toy model, we estimate that faint, low-$M_{\bigstar}$ ($\lesssim10^{8.5}\,M_{\odot}$) galaxies spend only $\approx10^{+4}_{-3}\%$ of their time in a U13151-like phase. However, depending on $f{\rm esc}$, these episodes could contribute substantially to the escaping ionizing photons from the low-mass galaxy population. This suggests that such short-lived episodes of intense star formation could substantially enhance the ionizing photon output of low-mass galaxies, potentially playing an important role during Cosmic Reionization (see also \citealt{jecmen_glimpse_2026}).

\end{itemize}

The advent of JWST has undoubtedly changed the way we study the early Universe. In its first years of operation, it has begun to fulfil its promise of finding the very first galaxies, pushing the redshift frontier into epochs that were previously beyond our reach through the discovery of remarkably bright and distant systems (e.g., \citealt{carniani_spectroscopic_2024, naidu_cosmic_2026}). At the same time, especially thanks to gravitational lensing, JWST is also allowing us to witness perhaps the first instants of galaxy assembly in unprecedented detail.  
The discovery of such building blocks, including U13151, the Cosmic Gems Arc (\citealt{adamo_bound_2024, messa_jwst_2026}), the Misty Moons (\citealt{nakane_venus_2025}), and other similar systems, offers a rare glimpse into the earliest phases of galaxy formation.  

In this work, we show that detailed rest-frame UV-to-optical studies of such ultra-faint, ultra-low-$M_{\bigstar}$ systems are now possible even when the Universe was only $\approx500$ Myr old, with MIRI playing a crucial role in extending our view beyond the rest-frame UV at such high redshifts. Our results also highlight that, to shed light on the (perhaps) first moments of galaxy formation, we must reach the intrinsically faintest systems. If it true that U13151-like galaxies at much higher redshifts could represent early stages of the observed population of massive and luminous Blue Monsters at $z>10$, then strong gravitational lensing offers the clearest, and easiest, path toward revealing and characterizing these otherwise inaccessible and observationally very time-demanding sources. In this respect, the VENUS campaign (PID: 6882; PIs: S. Fujimoto \& D. Coe) represents an ideal starting point. Looking ahead, both Euclid and the Roman Space Telescope promise to uncover many more strong gravitational lenses, opening the way to much larger samples of these systems. Step by step, discoveries of new U13151-like objects will allow us to piece together the earliest chapters of stellar-mass assembly and, ultimately, to understand how the first galaxies grew into the objects that populate the present-day Universe.

\begin{acknowledgments}
The authors thank Massimo Stiavelli, Joris Witstok, Sandro Tacchella, Steven Finkelstein, Andrea Ferrara, and Julian Muñoz
for useful discussions, Elka Rusta for sharing their {\sc NEFERTITI} simulations (\citealt{rusta_linking_2024}), which enriched our discussion, and Mario Llerena for providing the data points used in Figure~\ref{fig:xion_vs_z}.

\vspace{3mm}

This work is based on observations made with the NASA/ESA/CSA James Webb Space Telescope, obtained at the Space Telescope Science Institute, which is operated by the Association of Universities for Research in Astronomy, Incorporated, under NASA contract NAS5-03127. The JWST data presented in this article were obtained from the Mikulski Archive for Space Telescopes (MAST) at the Space Telescope Science Institute. Support for Program number \#5578 and \#8051 was provided through a grant from the STScI under NASA contract NAS5-03127.

\vspace{3mm}
PR, VK, and TYYH acknowledge support from the University of Texas at Austin Cosmic Frontier Center. 

Support for program \#8051 and \#5578  was provided by NASA through a grant from the Space Telescope Science Institute, which is operated by the Association of Universities for Research in Astronomy, Inc., under NASA contract NAS 5-03127. PR acknowledges support from NASA through STScI award JWST-GO-8051 and JWST-GO-5578.

JAM, CPJ, AVO acknowledge support by grant PID2024-158856NA-I00 from the Spanish Ministry of Science and Innovation/State Agency of Research MCIN/AEI/10.13039/501100011033 and by “ERDF A way of making Europe”. JAM, LC, PGPG, CPJ, AVO acknowledge support by grant CSIC/BILATERALES2025/BIJSP25022.
PS acknowledge financial support from INAF RF2024 Large Grant "UNDUST: UNveiling the Dawn of the Universe with JWST". MC acknowledges INAF GO Grant 2024 ”Revealing the nature of bright galaxies at cosmic dawn with deep JWST spectroscopy”. KIC acknowledges funding from the Dutch Research Council (NWO) through the award of the Vici Grant VI.C.212.036. JMH acknowledges support from the Evolving Universe Fellowship, which is made possible by a generous donation from Dr. Keiko Miwa Ross; JMH also acknowledges support from JWST Program \#8544. A.V.-O. is supported by the European Research Council (ERC) under the European Union’s Horizon 2020 research and innovation programme (DistantDust, Grant agreement No. 101117541).

\end{acknowledgments}

\bibliography{references}{}
\bibliographystyle{aasjournalv7}



\end{document}